\documentclass[a4paper,twoside]{article}
\newcommand{\affil}[1]{$^{\rm #1}$}
\usepackage[authoryear]{natbib}
\bibpunct{(}{)}{;}{a}{}{,}
\usepackage{graphicx}
\date{} 
\title{\large\bf\flushleft  Origin 
of structural and kinematical properties of the Small Magellanic Cloud}
\author{\parbox{\textwidth}{\flushleft
\vspace{-0.5cm}
{\it Kenji Bekki\affil{A,C} and Masashi Chiba\affil{B}}\\
\vspace{0.4cm}
{\small \affil{A}\, 
School of Physics, University of New South Wales, Sydney 2052, Australia}\\
{\small \affil{B}\,Astronomical Institute,
Tohoku University, Sendai, 980-8578, Japan}\\
{\small \affil{C}\,Email: bekki@phys.unsw.edu.au}}}
\begin{document}
\twocolumn[
\begin{changemargin}{.8cm}{.5cm}
\begin{minipage}{.9\textwidth}
\vspace{-1cm}
\maketitle
%
%
\small{\bf Abstract:}

We investigate structural, kinematical, and chemical properties of stars
and
gas
in the Small Magellanic Cloud (SMC) interacting with the Large
Magellanic Cloud
(LMC) and the Galaxy based on a series of self-consistent
chemodynamical simulations.
We adopt a new ``dwarf spheroidal model'' in which the SMC initially
has both old stars
with a spherical spatial distribution and an extended HI gas disk.
We mainly investigate SMC's evolution for the last $\sim$ 3 Gyr within
which
the Magellanic stream (MS) and the Magellanic bridge (MB) can be
formed as a result of
the LMC-SMC-Galaxy interaction. Our principal results, which can be
tested against
observations, are as follows.
The final spatial distribution of the old stars projected onto the sky
is
spherical even after the strong LMC-SMC-Galaxy interaction, whereas that
of the
new ones is significantly flattened and appears to form a bar
structure. Old stars have
the line-of-sight velocity dispersion ($\sigma$) of $\sim$ 30
km~s$^{-1}$ and
slow rotation with the maximum rotational velocity ($V$) of less than
20 km~s$^{-1}$
and show asymmetry in the radial profiles.
New stars have  a smaller $\sigma$
than old ones  and a significant amount of rotation
($V/\sigma >1$).
HI gas shows
velocity dispersions of $ \sigma = 10-40$  km s$^{-1}$,
a high maximum rotational velocity  ($V \sim 50$ km s$^{-1}$),
and the spatial distribution similar to that of new stars.
The new stars with ages younger than 3 Gyr show a negative
metallicity gradient in the sense that  more metal-rich stars
are located in the inner regions of the SMC.
The MB inevitably contains  old stars
with the surface mass densities of $6-300 \times 10^4 {\rm M}_{\odot}$
deg$^{-2}$
depending on initial stellar distributions of the modeled SMC.
We find that the dwarf spheroidal model can explain more
self-consistently the
observed kinematical properties of stars and gas, compared with
another type of the
model (``the disk model'') in which the SMC initially consists of
stellar and gas disks.
We suggest that the SMC needs to be modeled as having
a spheroidal component rather than being a purely disk for better
understanding the SMC evolution. \\



\medskip
\medskip
\end{minipage}
\end{changemargin}
]
\small

\section{Introduction}

Dynamical interaction between the SMC, the LMC, and the Galaxy
have long been considered to be closely associated
not only with the evolution  of the SMC 
(e.g., Yoshizawa \& Noguchi 2003; YN03)
but also with the formation of the MS (e.g., Putman et al. 1998)
and the MB (e.g., Muller et al. 2004).
The observed structural, kinematical, and chemical
properties of stars, star clusters,  and gas 
in the SMC  have been extensively discussed
in variously different aspects of the SMC,
such as the star formation history 
(e.g., Russell \& Dopita 1992;  Pagel \& Tautvaisiene 1998; Piatti et al. 2007),
the total dynamical mass (e.g.,  Dopita et al. 1985; 
Stanimirovi\'c et al.  2004, S04),
and the interaction history with the LMC
(YN03; Bekki \& Chiba 2007).
One of the key, longstanding problems related to the SMC as a galaxy
is to understand how the past dynamical interaction between
the SMC, the LMC, and the galaxy could have 
influenced the chemical and dynamical evolution and the star formation
history of the SMC (e.g., Westerlund 1997).

In order to solve the above mentioned key problem,
we need to know not only three-dimensional (3D) space motions
of the Magellanic Clouds (MCs) but also dynamical and chemical
properties of stars with different ages and metallicities.
Recent proper motion measurements of the MCs and numerical
studies of the orbits of the MCs based on the proper motion measurements
have suggested that the MCs can be  a unbound with each other
and they are unbound to the Galaxy (Kallivayalil et al. 2006, K06;
Besla et al. 2007).
The two-dimensional (2D) distributions of stellar populations
with different ages and metallicities derived from AGB stars
have revealed star formation histories in different local regions
of the SMC
(Cioni et al. 2006).
Observational studies of stellar populations
of the SMC  based on their color-magnitude diagrams
have revealed  not only the 2D structures dependent on ages
and metallicities but also a more precise global star formation
history of the SMC (e.g., Harris \& Zaritsky  2004).
Thus, these observations on 2D structures
of different stellar populations, 
if compared with the corresponding
theoretical studies,
can provide a vital clue to the above-mentioned key question
of the SMC.

Previous theoretical models, however, did not provide
useful predictions on the 2D (or 3D) structures 
of stars with different ages and metallicities
owing to the strong limitations in their numerical simulations.
YN03 discussed the star formation history of the SMC
and the origin of structural and kinematical properties
of the MS.
They however did not discuss chemical and dynamical
properties of stars and gas {\it within the SMC}.
Although previous one-zone chemical evolution 
models discussed the long-term chemical evolution
history of the SMC (e.g., Spite et al. 1988; 
Tsujimoto et al. 1995; Idiart et al. 2007),
they did not  discuss dynamical properties of the SMC.
Thus no theoretical studies have been done which attempt to investigate
{\it both dynamical and chemical properties}  of the SMC
and thereby provide reasonable explanations for
the above-mentioned observations.
Given that only chemodynamical simulations can provide
predictions on 3D-dynamical and chemical properties
in a self-consistent manner,
it is vital for theoretical studies
to investigate the above-mentioned key observational
properties 
based on chemodynamical simulations of the SMC.

The purpose of this paper is thus to 
numerically investigate physical properties
of the SMC in a fully self-consistent manner and thereby
try to provide answers for the above-mentioned unresolved problems.
We mainly discuss the origin of structural, kinematical,
and chemical properties of the SMC's stellar and gaseous
components derived  in previous observations 
(e.g., Suntzeff et al. 1986;  Torres \& Caranza 1987;
Hardy et al. 1989; Russell \& Dopita 1992; Hatzidimitiriou et al. 1993;
S04).
We also discuss the latest observations 
that previous theoretical works did not discuss:  the structure of
the stellar halo (No\"el \& Gallart 2007),
kinematical properties of older stellar populations (e.g.,  Harris \& Zarisky 2006),
possibly purely gaseous streams  in the MB (Harris  2007),
and bifurcated structures and kinematics of the MC 
(Matthews \& Staveley-Smith 2007).
We however do not intend to discuss some important aspects
of the SMC evolution,  
such as the long-term evolution of the MCs
and the formation of globular cluster (GCs),
which were discussed 
to some extent in our previous papers (Bekki et al. 2004;
Bekki \& Chiba 2005; paper I, Bekki 2007) in the preset study. 
We plan to discuss these in our paper III (Bekki \& Chiba 2008a)
based on more sophisticated
and self-consistent models not only for the MCs but also for
the Galaxy.

The layout of this paper is as follows. In \S 2, we summarize our numerical
models used in the present study and describe the methods for analyzing
structure and kinematics of the simulated SMC. In \S 3, we present numerical
results on the time evolution of morphology, metallicity distribution, and
dynamical properties of the SMC. In \S 4, we discuss the above-mentioned   outstanding
issues related to formation and evolution of the SMC. 
The conclusions of the preset study are given in \S 5. 
Since the origin of the MS has been already discussed extensively
in previous numerical simulations
(e.g.,  Gardiner \& Noguchi 1996, GN;
Connors et al. 2006, C06), we briefly discuss it in the present paper.
The present paper is still based on similar approaches
used in previous numerical works (e.g., simple analytic models for dynamical
friction and the fixed isothermal density profile for the Galactic
dark matter halo).
The more sophisticated models in which the LMC, the SMC,
and the Galaxy are represented by fully consistent N-body models
will be described in our forthcoming papers (Bekki \& Chiba 2008a).

\begin{figure}[h]
\begin{center}
\includegraphics[scale=0.6, angle=0]{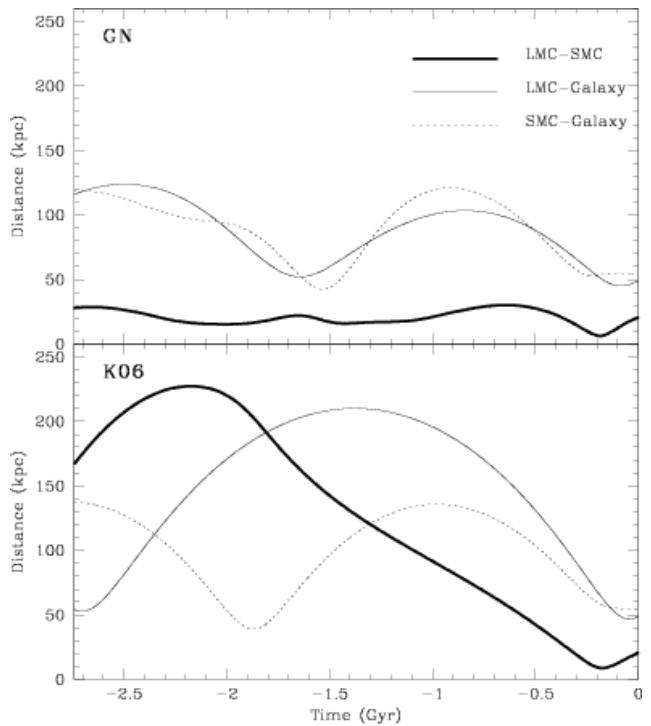}
\caption
{
Orbital evolution of the MCs for the
models C1 and C2 with  
the GN (upper) and the K06 (lower)
velocity types.
The time evolution of distances between MCs
(thick solid), the LMC and the Galaxy (thin solid),
and the SMC and the Galaxy (dotted) are separately shown.
Here $M_{\rm LMC}=2.0 \times 10^{10} {\rm M}_{\odot}$
and $M_{\rm SMC}=3.0 \times 10^{9} {\rm M}_{\odot}$
are assumed for the two models.
}
\label{figexample}
\end{center}
\end{figure}

\section{Model}

The present investigation is two-fold.
First, we investigate orbital evolution of the MCs 
with respect to the Galaxy 
for a given set of initial velocities of the MCs
by using a backward integration
scheme (Murai \& Fujimoto 1980; MF).
Then we investigate chemodynamical evolution of the SMC on the
orbits  by using  our original  
chemodynamical simulations with star formation models.
Since the details of the methods for orbital calculations
and numerical techniques for chemodynamical simulations
of the MCs are given in the paper I,
we briefly describe these in the present paper.
The present model is significantly  better than the previous
ones (e.g., GN and C06) in the sense that
(i) the SMC model is much more consistent with
the observations and (ii) star formation and chemical evolution
processes are better modeled.

\subsection{Derivation of the orbits of the MCs}

The gravitational potential of the Galaxy ${\Phi}_{\rm G}$
is assumed to have the logarithmic potential;
\begin{equation}
{\Phi}_{\rm G}(r)=-{V_0}^2 \ln r ,
\end{equation}
where  $r$ and $V_{0}$ are the distance from the Galactic center
and the constant rotational velocity (= 220 km s$^{-1}$), respectively. 
We consider this  potential, firstly because the present model
can be consistent with that of the paper I (and those of previous
papers, e.g., MF) and secondly because we can estimate the
drag force due to  dynamical friction between the MCs and the Galaxy
in an  analytical way for the orbital calculations of the MCs.
Thanks to this adoption of the potential,
we can compare the present results with previous ones
(e.g., GN, C06,  K06, and Besla et al. 2007).

Owing to the adopted isothermal density profile of the halo
(i.e., ever increasing mass with radius),
the MCs are more likely to show bound orbits for  given
initial velocities.  If we adopt that the latest model
for the Galaxy with a more realistic
halo model (e.g., Widrow \& Dubinski 2005), the orbital 
evolution of the MCs can be different owing to the smaller mass
of the Galaxy in its outer part ($R>100$ kpc).
However,  it is not practical  for the present study to
adopt the model by Widrow \& Dubinski (2005),
because analytical estimation of the drag force due to dynamical
friction between the MCs and the Galaxy is not feasible.

The LMC is assumed to have the Plummer potential;
\begin{equation}
 {\Phi}_{\rm L}(r_{\rm L})=-M_{\rm LMC}/{({r_{\rm L}}^2+a_{\rm L}^2)}^{0.5},
\end{equation}
where $M_{\rm LMC}$, $r_{\rm L}$, and $a_{\rm L}$ are 
the total mass of the LMC, the distance from the LMC, and
the effective radius, respectively. We adopt the same value
of  $a_{\rm L}$ (=3 kpc) as previous numerical studies
adopted (e.g., GN).
The SMC is also assumed to have the Plummer potential;
\begin{equation}
 {\Phi}_{\rm S}(r_{\rm S})=-GM_{\rm SMC}/{({r_{\rm S}}^2+a_{\rm S}^2)}^{0.5},
\end{equation}
where $M_{\rm SMC}$, $r_{\rm S}$, and $a_{\rm S}$ are 
the total mass of the SMC, the distance from the SMC, and
the effective radius, respectively. We adopt the same value
of  $a_{\rm S}$ (=2 kpc) 
as previous numerical studies
adopted (e.g., GN).

 We consider the dynamical friction due to the presence of the Galactic
dark matter halo both for the LMC-Galaxy interaction
and for the SMC-Galaxy one and adopt the 
following expression (Binney \& Tremaine 1987);
\begin{equation}
F_{\rm fric, G}=-0.428\ln {\Lambda}_{\rm G} \frac{GM^2}{r^2},
\end{equation}
where $r$ is the distance of the LMC (the SMC)
from the center of the Galaxy.
The mass $M$ is either $M_{\rm LMC}$ or $M_{\rm SMC}$,
depending on which Cloud's orbit (i.e., LMC or SMC) we calculate.
We adopt the reasonable value of 3.0 
for the Coulomb logarithm ${\Lambda}_{\rm G}$
(GSF; GN)
both in  the orbital calculation of the LMC and in that of the SMC.

By integrating equations of the motions of the MCs 
toward the past from the present epoch,
we investigate orbital evolution of the MCs 
for given $M_{\rm LMC}$, $M_{\rm SMC}$,
initial positions  and velocities of the MC.
We adopt the reasonable sets of orbital parameters that are 
consistent with observations.
The current Galactic coordinate $(b,l)$, where $l$ and $b$ are
the Galactic longitude and latitude, respectively, is $(-32.89, 280.46)$
for the LMC and $(-44.30,302.79)$ for the SMC.
Therefore the current positions $(X,Y,Z)$ of the LMC, 
$(X_{\rm L},Y_{\rm L},Z_{\rm L})$
in units of kpc, 
are $(-1.0,-40.8,-26.8)$ and those of
the SMC, $(X_{\rm S},Y_{\rm S},Z_{\rm S})$,
are $(13.6,-34.3,-39.8)$.
The current distance and the Galactocentric radial velocity of the LMC (SMC)
is 80 (7) km s$^{-1}$.

The current space velocities 
or  $(U,V,W)$ in units of km s$^{-1}$ are the most important
parameters that determine the orbital evolution
of the MCs in the present models.
They  are represented by $(U_{\rm L},V_{\rm L},W_{\rm L})$
for the LMC and  by $(U_{\rm S},V_{\rm S},W_{\rm S})$ for the SMC.
We mainly investigate the following two velocity types:
The ``GN type''
with $(U_{\rm L},V_{\rm L},W_{\rm L})$ = $(-5,-225,194)$
and $(U_{\rm S},V_{\rm S},W_{\rm S})$  = $(40,-185,171)$
and the ``K06 type''
with $(U_{\rm L},V_{\rm L},W_{\rm L})$ = $(-86,-268,252)$
and $(U_{\rm S},V_{\rm S},W_{\rm S})$  = $(-87,-247,149)$.
The former is the same as those used in the models by GN
whereas the latter is consistent with observations by K06.

Fig. 1 describes the past $\sim$ 3 Gyr orbital evolution of the MCs 
in the models with the GN and K06 velocity types 
for which $M_{\rm LMC}=2 \times 10^{10} {\rm M}_{\odot}$
and $M_{\rm SMC}=3 \times 10^{9} {\rm M}_{\odot}$
are assumed. 
Here  negative values of the time, $T$, represent the past, with
$T=0$ corresponding to the present epoch. 
As shown in this figure,
the present orbital period of the MCs 
about the Galaxy is $\sim$ 1.5 Gyr for the adopted
gravitational potential and the masses of the MCs.
The  strong LMC-SMC interaction around 1.5 Gyr can be
seen only in the model with the GN type, which suggests
that the MS can not be formed from the SMC in the model
with the K06 type.

For a given set of initial positions and velocities of the MCs
predicted by the above backward integration scheme,
we dynamically evolve the MCs by using GRAPE-SPH simulations
(i.e., integrating the equation of motion {\it forward})
for a fixed potential of the Galaxy.
Since the mass of the SMC is very small in comparison
with the Galaxy, we consider that the tidal effects of the Galaxy
on the SMC (e.g., stripping of stars from the SMC) can be 
properly treated by  our models.
Dynamical evolution of the SMC under the {\it live potential}
of the Galaxy will be investigated by our future studies 
and the results will be compared with those of the present study. 

We mainly discuss the models with
 ``classical'' bound orbits of the LMC and the SMC (i.e., GN-type)
around the Galaxy in order to reproduce well the observed properties
of the MS.  The bound orbits are also suggested by
Shattow \& Loeb (2008) which adopted $V_0$ a factor of 14\% larger
than 220 km s$^{-1}$ adopted in the present study.
They also suggested that the bound orbits can be consistent
with the latest proper motion measurements by K05 as long as
the higher circular velocity is adopted. 
Thus our main investigation of the GN-type bound orbits
can be justified in the present study.

\begin{figure}[h]
\begin{center}
\includegraphics[scale=0.4, angle=0]{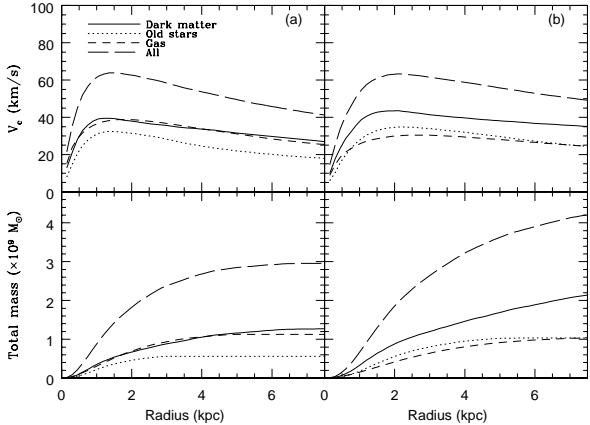}
\caption
{
Circular velocities ($V_{\rm c}$) dependent on the distance ($R$)
from the center of the SMC (upper) and total masses within $R$
(lower) for the model G5
with $M_{t}=3.0 \times 10^9 {\rm M}_{\odot}$
(a) and G1 with $M_{t}=4.5 \times 10^9 {\rm M}_{\odot}$ (b).
$V_{\rm c}$ is calculated  separately for the mass profiles
of the dark matter halo (solid), the old stars (dotted),
the gas (short-dash), and  all components (long-dash).
}
\label{figexample}
\end{center}
\end{figure}

\begin{figure}[h]
\begin{center}
\includegraphics[scale=0.5, angle=0]{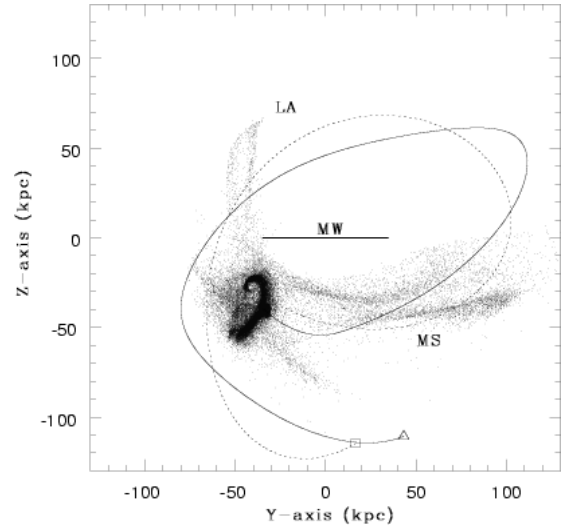}
\caption
{
The final distribution of gas projected onto the $y$-$z$ plane
for the model C1. The orbits of the LMC (triangle) and the SMC (square)
with respect to the center of the Galaxy are shown by solid and
dotted lines, respectively. For comparison,
the Galaxy with the disk size of 17.5 kpc is shown by a thick solid
line.
}
\label{figexample}
\end{center}
\end{figure}

\begin{table*}
\centering
\begin{minipage}{175mm}
\caption{Model parameters for N-body simulations}
\begin{tabular}{cccccccccc}
{Model name
\footnote{``G'' represents  gaseous models
in which gas physics, star formation
in gas, and chemical enrichment processes
are all included whereas ``C'' represents collisionless models
in which such processes are switched off. 
``I'' represents isolated models in which interaction between
the MCs and the Galaxy is not included at all.}}
& {$M_{\rm SMC}$
\footnote{The total mass of the SMC  in units of $10^{10}$  ${\rm M_{\odot}}$.}}
& {Velocity types 
\footnote{``GN'' and ``K06'' describes the orbits of the MCs
that are consistent with those by GN and K06, respectively.}}
& {Morphology
\footnote{The initial morphology of old stars: ``dSph'' and ``disk''
represent the dwarf spheroidal and the disk models, respectively.
We use the term of dwarf spheroid for convenience, though it would not be so
appropriate for the SMC to be classified  as a ``dwarf'' spheroid
for its luminosity.  }}
& {$R_{\rm SMC}$
\footnote{The size of the SMC which corresponds to the tidal radius
in units of kpc.}}
& {$r_{\rm s}$
\footnote{The ratio of the stellar disk size ($R_{\rm SMC,s}$)
to the SMC one ($R_{\rm SMC}$).}}
& {$f_{\rm g}$
\footnote{The mass ratio of gas to stars in the SMC.}}
& {$r_{\rm g}$
\footnote{The size ratio  of the gaseous disk to the stellar spheroid or disk
in the SMC.}}
& {$T_{\rm i}$
\footnote{The initial gaseous temperate in units of K.
The low temperature of cold gas is adopted so that the outer disk
can keep its initial thin configuration for most models.}}
& Comments \\
G1 &  0.45  &  GN  & dSph & 7.5 & 0.5 & 1.0  &  2.0  & 250 & fiducial  \\
G2 &  0.45  &  GN  & dSph & 7.5 & 0.25 & 1.0  &  2.0  & 250 & more compact \\
G3 &  0.45  &  GN  & disk & 7.5 & 0.25 & 1.0  &  4.0  & 250 & disk model  \\
G4 &  0.45  &  GN  & dSph & 7.5 & 0.5 & 1.0  &  2.0  & 250 & $y_{\rm met}=0.002$ \\
G5 &  0.30  &  GN  & dSph & 5.0 & 0.5 & 2.0  &  2.0  & 250 & lower mass  \\
G6 &  0.30  &  GN  & dSph & 7.5 & 0.25 & 1.0  &  2.0  &  250 & \\
G7 &  0.45  &  K06  & dSph & 7.5 & 0.5 & 1.0  &  2.0  & 250 & orbit from K06\\
I1 &  0.45  &  $-$  & dSph & 7.5 & 0.5 & 1.0  &  2.0  & 250 & isolated   \\
I2 &  0.45  &  $-$  & dSph & 7.5 & 0.5 & 1.0  &  2.0  & $10^3$ &   \\
I3 &  0.45  &  $-$  & dSph & 7.5 & 0.5 & 1.0  &  2.0  & $10^4$  &    \\
C1 &  0.30  &  GN  & dSph & 7.5 & 0.25 & 2.0  &  2.0  & 250 & collisionless   \\
C2 &  0.30  &  K06  & dSph & 7.5 & 0.25 & 2.0  &  2.0  & 250 &   \\
C3 &  0.45  &  GN  & dSph & 7.5 & 0.5 & 1.0  &  2.0  & 250 &  \\
C4 &  0.30  &  GN  & dSph & 5.0 & 0.5 & 2.0  &  2.0  & 250 &  \\
\end{tabular}
\end{minipage}
\end{table*}

\subsection{GRAPE-SPH chemodynamical simulations}

For a given orbit of the SMC,
we investigate chemodynamical evolutions of the SMC 
by adopting the  following two different galaxy models: 
(i) the ``dwarf spheroidal'' model
in which the SMC has a stellar spheroidal and an extended HI gas disk 
and (ii) the ``disk'' one in which the SMC has stellar and gaseous disks.
Although the SMC is classified as an irregular dwarf of DDO type Ir IV-V
(van den Bergh 2000), 
the usage of ``dwarf'' spheroidal for the SMC
would not be so appropriate,
given that the SMC's stellar mass is as large as $10^9 {\rm M}_{\odot}$
(S04): we use this term just for convenience.

Although the above disk model was adopted in 
previous models of the SMC (MK, GN, YN, and C06),
we mainly investigate the dwarf spheroidal model in the present study.
This is mainly because the dwarf spheroidal model can better explain
the observed structural and kinematical properties of the SMC.
The LMC is represented by a test particle with the mass
of $M_{\rm LMC}$ and the Plummer potential described in \S 2.1
for most models.
The results of simulations in which both the LMC and the SMC
are represented by  gaseous and stellar particles in a fully
self-consistent manner will be described in (Bekki \& Chiba 2008a) 
with main points of analysis significantly
different from those of the present one.

\subsubsection{The dwarf spheroidal model}

A significant fraction of
dwarf elliptical and spheroidal galaxies  are observed to have exponential
luminosity profiles
(e.g., Ichikawa, Wakamatsu, \& Okamura 1986
Andredakis \& Sanders 1994): we  do not discriminate dwarf ellipticals
and spheroids just for convenience in the present paper.
Therefore the projected radial density profile (${\Sigma}_{\rm sph}$) of
the stellar spheroid with the mass of $M_{\rm sph}$
and the scale length of $a_{\rm sph}$ is
assumed to be an exponential one:
\begin{equation}
{\Sigma}_{\rm sph} (R) = {\Sigma}_{\rm sph,0} \times \exp (-R/a_{\rm sph}),
\end{equation}
where ${\Sigma}_{\rm sph,0}$ and $R$ are the central surface density
and the projected radius from the center of the spheroid
(i.e., the center of the SMC), respectively.
In order to derive the three dimensional (3D) density field
(${\rho}_{\rm sph} (r)$, where $r$ is the distance from the
center of the spheroid)
from ${\Sigma}_{\rm sph}$, we can use the following formula
(Binney \& Tremaine  1986):
\begin{equation}
{\rho}_{\rm sph} (r)
 = -\frac{1}{\pi} {\int}_{r}^{\infty}
 \frac{d{\Sigma}_{\rm sph}(R)}{dR}
 \frac{dR}{\sqrt{R^2-r^2}}.
\end{equation}

We numerically estimate the ${\rho}_{\rm sph} (r)$ profile
for a given ${\Sigma}_{\rm sph} (R)$ in a model with
$M_{\rm sph}$ and $a_{\rm sph}$.
The outer cut-off radius of the stellar spheroid ($R_{\rm SMC, s}$)
is assumed to be  a  free parameter. 
The spheroid is assumed to be embedded in a massive dark matter
halo with the total mass of $M_{\rm dm}$
and the truncation radius of $R_{\rm SMC}$.
This  $R_{\rm SMC}$ represents the size of the SMC within which
almost all particles are included.
The radial density profile
is assumed to have
either the universal ``NFW'' one  with the central cusp 
(Navarro, Frenk \& White 1996)
or a flat  ``core''   (Burkert 1995; Salucci \& Burkert 2000; 
referred to as the SB profile
from now on).
The NFW profile is described as:
\begin{equation}
{\rho}_{\rm dm}(r)=\frac{\rho_{0}}{(r/r_{\rm s})(1+r/r_{\rm s})^2},
\end{equation}
where $r$,  $\rho_{0}$,  and $r_{\rm s}$ are the distance from the center
of the cluster, the scale density, and the scale-length of the dark halo,
respectively. 
The SB profile is described as:
\begin{equation}
{\rho}_{\rm dm}(r)=\frac{\rho_{dm,0}}{(r+a_{\rm dm})(r^2+{a_{\rm dm}}^2)},
\end{equation}
where $\rho_{dm,0}$ and $a_{\rm dm}$ are the central dark matter
density and the core (scale) radius, respectively.
The halo is truncated at 7.5 kpc from the center of the SMC.

S04 demonstrated that (i) the HI rotation curve is very slowly rising
within the central few kpc of the SMC 
with the maximum circular velocity of $\sim 60$ km s$^{-1}$ 
and (ii) the total mass of the SMC is $\sim 2.4 \times 10^{9} {\rm M}_{\odot}$ 
within the central $\sim 3$ kpc.
These observed properties  of the SMC
are  better explained by the SB profile (see more detailed discussions  on this for 
Bekki \& Stanimirovi\'c 2008): the NFW profile with the reasonable parameters 
for the SMC (e.g., c=$5-20$,
where $c$ is the ratio of the virial radius to  $r_{\rm s}$)
is not so consistent with the above observation (i).
We find that the SB profiles with $a_{\rm dm}$ as large as $5-7$ kpc
can be more consistent with the observations
for  $M_{\rm SMC}=3-4.5 \times 10^9 {\rm M}_{\odot}$,
though the slowly rising profile can not be exactly the same
as the observed.
We therefore 
show the results of the models with the SB profiles. 
We assume that $M_{\rm dm}/M_{\rm sph}$ is 2.33
and the halo is truncated at the tidal radius ($r_{\rm t}$)
of the SMC.
The previous models for the SMC  assumed that
$r_{\rm t}$ ranges  from  5 kpc (GN)
to 7.5 kpc (C06). 
$R_{\rm SMC}$ is the same as $r_{\rm t}$ in the present model
and regarded as a free parameter.

The stellar spheroid is assumed to 
be supported purely by velocity dispersion
and its dispersion is assumed to be isotropic.
We therefore estimate the velocities of  old stellar  particles
from the gravitational potential
at the positions where they are located.
In detail,
we first calculate the one-dimensional isotropic dispersion according to
the following formula:
\begin{equation}
{\sigma}^{2}(r)=-\frac{U(r)}{3},
\end{equation}
where $U(r)$ is the gravitational potential at the position $r$
which is determined by the mass distributions of the dark matter
halo and the stellar spheroid.
Then we allocate a velocity to  stellar  particle
so that the distribution of velocities
of these particles can have a Gaussian form with a dispersion equal
to ${\sigma}^{2}(r)$.
We adopt this method as an approximation
and thus need to let the halo dynamically relaxed 
to obtain  a real  dynamically equilibrium 
for a given  distribution of the SMC:
this approximation is quite useful, in particular,
for implementing the triaxial potential of the  disk model 
later described.

The SMC is assumed to have an extended HI gas disk with
the total gas mass of $M_{\rm SMC,g}$,
the size of $R_{\rm SMC, g}$,
and 
the exponential profile for the projected gaseous density.
The scale length of the disk is set to be $0.2 R_{\rm SMC, g}$
for all spheroidal models.
The size ratio ($r_{\rm g}$) of  $R_{\rm SMC, g}$ to $R_{\rm SMC, s}$
is a key parameter which determines whether the MS is composed
almost purely of gas (YN03).
Furthermore  HI diameters of gas-rich galaxies are generally observed to be 
significantly larger
than their optical disks (Broeils \& van Woerden 1994).
Guided by these previous works,
we investigate models with $2\le r_{\rm g} \le 4$
in the present study.
The mass ratio ($f_{\rm g}$) 
of $M_{\rm SMC,g}$ to $M_{\rm SMC,s}$ is a key parameter which
determines the masses of the MS and the LA (YN03).
We find that  $f_{\rm g}$ should be at least
$\sim 2.0$ for $M_{\rm SMC}=3 \times 10^9 {\rm M}_{\odot}$
and $\sim 1.0$  for $M_{\rm SMC}=4.5 \times 10^9 {\rm M}_{\odot}$ 
to explain the total gas mass in the MB and the MS regions.
We thus show the results of the models with $1 \le f_{\rm g} \le 2$
in the present study, though we have run models with  $f_{\rm g} \le 0.5$ 
as done in GN and C06.

Since the disk component in this spheroidal model
is composed purely of gas,
it is highly likely to be dynamically supported by rotation:
gas dynamics (e.g., cooling and shock)
in the disk can quickly dissipate away kinematic
energy resulting from random motion of gas.
Such a cold gas disk, however, can be quite unstable
against dynamical instability (e.g., ring or bar instability),
if the mass fraction
of the disk is significant (e.g., Binney \& Tremaine 1987).
For such a massive gas disk,
the disk may well have a significant amount of kinetic energy
due to random motion of the gas.
In addition to the rotational velocity made by the gravitational
field of the SMC, the initial radial and azimuthal velocity
dispersion are thus given to the gaseous  component according
to the epicyclic theory with Toomre's parameter 
(Binney \& Tremaine 1987) $Q$ = 1.5: thus the cold gas
is stabilized  by its random motion.

Since the SMC has a massive gaseous disk surrounding old stars,
we need to adopt the following unique procedures to construct
a dynamical equilibrium of the SMC. 
Firstly we  give velocities of dark matter particles for given
density profiles of the dark matter halo and the stellar spheroid.
Secondly we  let the dark matter halo evolve for ten dynamical time
scales with the spheroid being fixed.  
During this relaxation
process, the mass distribution of the halo can slightly change
so that the system can reach a new dynamical equilibrium.
Thirdly we allocate velocities for stellar particles 
according to the equation (9) and add a gas disk
to the system.  Fourthly, we again let the system evolve 
for ten dynamical time scales with the gas disk being fixed.
During this relaxation
process, the mass distribution of the spheroid (and the halo)  can slightly change
so that the system can reach a new dynamical equilibrium.
Finally, we give rotational velocities to gaseous particles
according to the circular velocities at their position 
for the new mass profile of the system.

Fig. 2 shows the rotation curves of two dwarf spheroidal  models
with  $M_{\rm SMC} =3.0 \times 10^9 {\rm M}_{\odot}$
and $R_{\rm SMC} =5.0$ kpc 
and with  $M_{\rm SMC} =4.5 \times 10^9 {\rm M}_{\odot}$
and $R_{\rm SMC} =7.5$ kpc.
These models  show the maximum circular velocities
of $\sim 60$ km s$^{-1}$,
which is consistent with HI kinematics of the SMC
(e.g., S04).
For models with small SMC masses 
($M_{\rm SMC} \sim 3.0 \times 10^9 {\rm M}_{\odot}$),
the initial SMC size should be as small as 5 kpc
so that the final total mass of the SMC
after efficient tidal stripping of stars and gas
can be more consistent with the observations by S04.
It should be stressed here that the rotation curves
in this small-mass model reach their peaks
at $\sim 1$ kpc and thus appear to be less consistent
with the observations by S04. 
The total particle numbers for collisionless and collisional
particles for the SMC are 100000 and 60000, respectively.
The total particle number of a MC system is  260000
(or larger, depending on whether the gas disk is modeled), when
the LMC is also represented by a fully consistent disk model
rather than a point-mass particle.
The total mass of the dark matter halo ($M_{\rm dm}$) in a model
with $M_{\rm SMC} = 4.5 \times  10^9 {\rm M}_{\odot}$
is about $2.4 \times 10^9 {\rm M}_{\odot}$:
$M_{\rm dm}$ depends on $f_{\rm g}$ and $M_{\rm SMC}$.

\begin{figure}[h]
\begin{center}
\includegraphics[scale=0.6, angle=0]{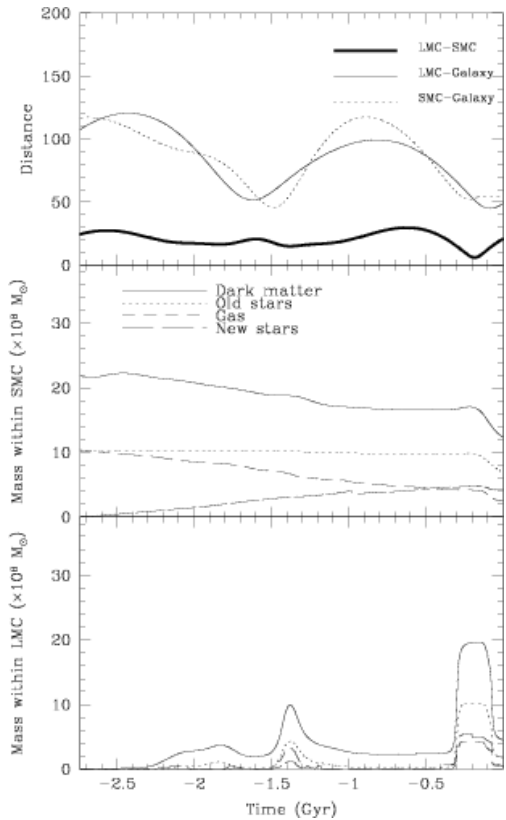}
\caption
{
The top panel shows the time evolution of the distances
between the LMC-SMC (thick solid), the LMC-Galaxy (thin solid),
and the SMC-Galaxy (dotted) for the fiducial model G1.
The middle panel shows the time evolution of
the total masses  within the central 5 kpc of the SMC for
the dark matter (solid),  old stars (dotted), gas (short-dash),
and new stars (long-dash).
The bottom panel shows the time evolution of the total masses
the dark matter (solid),  old stars (dotted), gas (short-dash),
and new stars (long-dash) that are initially in the SMC
and later in the central 7.5 kpc of the LMC at a give time step.
}
\label{figexample}
\end{center}
\end{figure}

\subsubsection{The disk model}

  The radial ($R$) and vertical ($Z$) density profile 
of the initially thin  disk of the SMC
in this disk model  are  assumed to be
proportional to $\exp (-R/R_{0}) $ with scale length $R_{0}$ = 375 pc 
and to  ${\rm sech}^2 (Z/Z_{0})$ with scale length $Z_{0}$ = $0.2R_{0}$
respectively.
The stellar disk is embedded in a massive dark matter halo
with the NFW radial density profile and the halo is truncated 
at the tidal radius of the SMC ($r_{\rm t}=7.5$ kpc). 
The mass ratio of the dark matter ($M_{\rm dm}$) to the stellar disk
($M_{\rm SMC, s}$) is set to be 2.33 for all disk models. 
In addition to the rotational velocity made by the gravitational
field of disk and halo component, the initial radial and azimuthal velocity
dispersion are given to the disk component according
to the epicyclic theory with Toomre's parameter 
(Binney \& Tremaine 1987) $Q$ = 1.5.
The vertical velocity dispersion at a given radius 
is set to be 0.5 times as large as
the radial velocity dispersion at that point.

The gaseous component in the disk model has the same properties
as that in the spheroidal model.
For consistency with the dwarf spheroidal models,
we investigate the disk models with $1 \le f_{\rm g}  \le 2$
and $2 \le r_{\rm g} \le 4$.
Thus, the initial structures and kinematics  of old stars
are quite different in the dwarf spheroidal and disk
models for the SMC  whereas those of the gas are the same
in the two models.
These initial differences in stellar components
can result in significant differences
in dynamical properties and star formation histories of the SMC
between the two different models.
The particle numbers for the three components 
in the disk models
are the same as those in the dwarf spheroidal model.

\subsubsection{Star formation and chemical enrichment}

Star formation
is modeled by converting  the collisional 
gas particles
into  collisionless new stellar particles according to the algorithm
of star formation  described below.
We adopt the Schmidt law (Schmidt 1959)
with exponent $\gamma$ = 1.5 (1.0  $ < $  $\gamma$
$ < $ 2.0, Kennicutt 1998) as the controlling
parameter of the rate of star formation.
The amount of gas 
consumed by star formation for each gas particle
in each time step 
is given as:
\begin{equation}
\dot{{\rho}_{\rm g}} = C_{\rm sf}  
{\rho_{\rm g}}^{\gamma},
\end{equation}
where $\rho_{\rm g}$ and $ C_{\rm sf}$ 
is the gas density around each gas particle 
and a  constant in the Schmidt law.
The  value of $ C_{\rm sf}$ is determined such that
the star formation rate in the isolated model for the Galaxy
with the disk mass of  $6 \times 10^{10}  {\rm M}_{\odot}$
and the gas mass fraction of 0.1
can be $\sim 1 {\rm M}_{\odot}$ yr$^{-1}$
(e.g., Bekki 1998).
We convert a gas particle into a field star only if the local
surface gas density  exceeds the observed threshold gas density
of $\sim$ 3 $M_{\odot}$ pc$^{-2}$ for 
Magellanic  dwarf irregular galaxies (Hunter et al. 1998).
We estimate the surface gas density of a SPH particle
from the volume gas density by assuming that the scale-height of the gas disk
in the SMC being 100 pc during simulations.
These stars formed from gas are called ``new stars'' (or ``young stars'')
whereas stars initially within a disk  are called ``old stars''
throughout this paper.

Chemical enrichment through star formation and supernova feedback
during the SMC evolution 
is assumed to proceed both locally and instantaneously in the present study.
We assign the metallicity of original
gas particle to  the new stellar particle and increase 
the metals of the each neighbor gas particle 
with the total number of neighbor gas particles equal to  $N_{\rm gas}$,
according to the following 
equation about the chemical enrichment:
  \begin{equation}
  \Delta M_{\rm Z} = \{ Z_{i}R_{\rm met}m_{\rm s}+(1.0-R_{\rm met})
 (1.0-Z_{i})m_{\rm s}y_{\rm met} \}/N_{\rm gas} 
  \end{equation}
here the $\Delta M_{\rm Z}$ represents the increase of metal for each
gas particle. $ Z_{i}$, $R_{\rm met}$, $m_{\rm s}$,
and $y_{\rm met}$  in the above equation represent
the metallicity of the new stellar particle 
(or that of original gas particle),
and the fraction of gas returned to interstellar medium,  the
mass of the new star, and the chemical yield, respectively.
The values of $R_{\rm met}$,  $y_{\rm met}$, 
and the initial metallicity  are set to
be 0.3 and 0.004, and 0.0032, respectively.
For these values, the final mean metallicity of
of the SMC 
is ${\rm [Fe/H]} = -0.56$: this value
is significantly smaller for a smaller $y_{\rm met}$.

Dynamical evolution of interstellar medium (ISM)
can be significantly influenced
by supernova explosion and the effects of supernova explosion
on ISM
depend on the  initial mass function (IMF) adopted in the present study.
For the stellar population formed in a GMC, we assume an IMF described
as $\psi (m_{\rm I}) = A{m_{\rm I}}^{-s}$, 
where $m_{\rm I}$ is the initial mass of
each individual star and the slope of $s=2.35$ corresponds to the Salpeter IMF.
The normalization factor $A$ is a function of $m_{\rm cl}$,
which is a total mass of a stellar system,
$m_{\rm l}$ (lower mass cut-off), and $m_{\rm u}$ (upper one):
\begin{equation}
A=\frac{m_{\rm cl} \times (2-s)}{{m_{\rm u}}^{2-s}-{m_{\rm l}}^{2-s}}.
\end{equation} 
 $m_{\rm u}$ is a parameter and
set to be $100 {\rm M}_{\odot}$ for most models
whereas $m_{\rm l}$ is regarded as a key free parameter
in the present study. We shows the results of
the canonical models with  $m_{\rm l}=0.1 {\rm M}_{\odot}$
and  $m_{\rm u}=100 {\rm M}_{\odot}$
and the top-heavy IMF ones with  $m_{\rm l}=1.0 {\rm M}_{\odot}$
and $m_{\rm u}=100 {\rm M}_{\odot}$.
The mass fraction of stars that can explode as type-II supernova
(i.e., $m_{\rm I} > 8 {\rm M}_{\odot}$) is 14\% for the canonical
model and 35\% for the top heavy IMF ones.
About 10 \% of the total energy of one supernova ($\sim 10^{51}$ erg)
is assumed to be  converted into kinematic energy of gas 
around the supernova (Thornton et al. 1998).

Some fraction of the  kinematic supernova energy ($E_{\rm SN,\it i}$)
of the $i$th stellar particle
is assumed to be returned
to gas particles (ISM) and then to change the kinetic energy
of gas particles around the $i$th stellar particle. Each $j$th
gas particle with the mass of $m_{g,j}$ around an $i$th stellar particle can receive
a velocity perturbation ($\Delta v_{\rm SN, \it j}$)
directed radially away from the $i$th stellar particle.
$\Delta v_{\rm SN, \it j}$  satisfies the relation:
 \begin{equation}
\displaystyle
f_{\rm SN} E_{\rm SN, \it i}= \frac{1}{2}
\sum_{j=1}^{N_{\rm nei, \it i}} m_{g,j}
{(\Delta v_{\rm SN, \it j})}^2,
\end{equation}
where $f_{\rm SN}$ represents the fraction of supernova energy
returned to ISM, $N_{\rm nei, \it i}$ is total number of
gas  particles that are located within the smoothing ($h_{\rm i}$) radius
around an $i$th stellar particle.
Although we investigate the models with 
$f_{\rm SN}=0$ (no feedback),
$f_{\rm SN}=0.1$,
and $f_{\rm SN}=1.0$ (maximum feedback),
we show the results of the models with the more realistic $f_{\rm SN}=0.1$
(Thornton et al. 1998).

\subsubsection{Orbital configurations}

The initial spin of a SMC (gas) disk in  a model is specified by two angles,
$\theta$ and $\phi$, where
$\theta$ is the angle between the $Z$-axis and the vector of
the angular momentum of a disk and
$\phi$ is the azimuthal angle measured from $X$-axis to
the projection of the angular momentum vector
of a disk onto the $X-Y$ plane.
Although we investigate many  models with different
$\theta$ and $\phi$,
we mainly describe the results of  the models with  
$\theta$ = 45$^{\circ}$ and $\phi$  = 230$^{\circ}$,
This is mainly because these models 
can not only explain fundamental properties of the MS (GN and YN)
but also reproduce dynamical properties of the present SMC.

\subsubsection{The GRAPE-SPH simulations}

We use the GRAPE (GRAvity pyPE) system (e.g., Sugimoto et al. 1991)
for the smoothed particle
hydrodynamics (SPH) 
in order to evolve self-gravitating stellar and gaseous systems.
Since we have already developed our original ``GRAPE-SPH'' code
capable of evolving
self-gravitating systems composed both of collisionless
and collisional masses (Bekki \& Chiba 2006),
we use the code to simulate the evolution of the MCs for the last
$\sim 3$ Gyr.
In this code,
we calculate (i) mutual gravitational interactions between all particles
and (ii) the list of neighbor particles for  each SPH gas particle
by using the GRAPE.
All other calculations related to hydrodynamics, star formation,
chemical evolution, and orbital evolution
of the MCs  are done by host computers for the GRAPE.

In order to implement SPH on the GRAPE system, 
we adopt numerical methods and techniques essentially the same
as those used for the TREESPH method (e.g., Hernquist \& Katz 1989, HK89).
We adopt the spherically symmetric spline kernel proposed by
Monaghan \& Lattanzio (1985) and determine the smoothing lengths
of SPH particles based on the numbers of their neighbor particles
(HK89). We use the same forms of artificial viscosity 
as those used in HK89 (given in equations 2.22 - 2.25 of HK89).
We adopt
an isothermal equation of state for the gas and determine
the proper range of the initial gas temperatures ($T_{\rm i}$) by running isolated
disk models.  We find that for the models
with $T_{\rm i}=10^4$ K, which is reasonable for luminous disk galaxies like
the Galaxy,  the initially thin gas disks quickly  become  very thick owing
to the higher gaseous pressure. A smaller $T_{\rm i}$
is necessary for the SMC with its smaller mass and, thus,  weaker self-gravity
in the vertical direction.
Although we investigate the models with
$T_{\rm i}$ =  250 K and $10^3$ K,
we mostly  describe  the
results of the models with $T_{\rm i}=250$ K.

We adopt the predictor-corrector algorithm that  is accurate
to  2nd order in time and space
in order to integrate the equations describing the time evolution
of self-gravitating systems
in the present GRAPE-SPH code.
The size of
the time step ($\delta t_{\rm i}$)
for each SPH particle is determined by the Courant
condition for the particle.
$\delta t_{\rm i}$ for a SPH particle
is derived from its gas density, sound velocity, velocity gradient,
and smoothing length according to the formula adopted in HK89. 
The minimum and  maximum  possible sizes of time steps for SPH particles
are set to be $1.67 \times 10^2$ yr and  $6.85 \times 10^5$ yr,
respectively.
We choose the above very small minimum possible size  so that
the time step size of every gas particle at a given time step can not be smaller
than the minimum possible size: otherwise the simulation needs to stop. 
The size of the time step for a collisionless particle 
(dark matter, old and new stars)
is fixed at $6.85 \times 10^5$ yr.

The gravitational softening lengths ($\epsilon$) are different between
collisionless and collisional particles and different between
models with different initial SMC sizes ($R_{\rm SMC}$). 
For the fiducial model (described later),
$\epsilon=1.70$ kpc for collisionless particles
and $\epsilon=199$ pc for collisional (gaseous) ones. 
Since our  main interest is on the 1kpc-scale structures and kinematics of the SMC,
we consider that 
the resolution of the present simulation is high enough to investigate them.

\subsubsection{Stellar population synthesis models}

We derive the two-dimensional (2D) distributions
of the projected $B$-band luminosity densities (${\Sigma}_{\rm L, B}$)
of the SMC based on the spatial distributions and the age- and 
metallicity-distributions of old and new stars.
In order to do this,
we use the predictions of stellar population synthesis
models by Vazdekis et al. (1996), which are 
available on his  website.
Since the observed
metallicity of the stellar component of the SMC
is with [Fe/H] = $-0.73$ $\pm$ 0.03 (Luck et al. 1998),
we use the $B$-band luminosities and stellar mass-to-light-ratios
for the  SSP models with ${\rm [Fe/H]} \sim -0.68$ and variously
different ages in Vazdekis et al. (1996).
We assume that the mean ages of old and new stars in the present simulation
are 5 Gyr and 0.5 Gyr, respectively, 
and thereby allocate the $B$-band luminosity to each stellar
particle.
We chose this 
set of stellar ages so that we can show how the SMC looks
on the sky if the SMC has different distributions of old and new
stars. Since our main focus in this paper is structural
and kinematical properties of the SMC,
we do not intend to discuss how the appearance of the SMC 
on the sky depends  on ages
of old and new stars.

We mainly investigate the projected 2D ${\Sigma}_{\rm L, B}$ distribution
of the present SMC 
for the region with  $0.1 {\rm h} < \alpha < 1.9 {\rm h}$
and $-77^{\circ} < \delta < -69^{\circ}$
(corresponding roughly  to the physical range of 4 kpc $\times$ 4 kpc)
so that observations can be compared with the simulations.
We divide the  region into $50 \times 50$ cells and thereby estimate
${\Sigma}_{\rm L, B}$ for each cell.
We then smooth out the 2D ${\Sigma}_{\rm L, B}$ field   by using
the Gaussian kernel with the size  0.5 times the cell size
so that we can  clearly see the {\it global}  2D ${\Sigma}_{\rm L, B}$ distribution
without unnecessarily dramatic  cell-to-cell variations  of ${\Sigma}_{\rm L, B}$ 
due to small numbers  of particles in some cells.

\subsection{Main points of analysis}

In the present GRAPE-SPH simulations, 
the SMC is modeled as a fully self-gravitating
N-body system whereas the LMC is represented by a test particle
with the above-mentioned Plummer potential for most models. 
The mass distribution of the Galaxy  is assumed to be  fixed 
so that dynamical influences of the MCs on the Galaxy
are not explicitly included in the models. 
We therefore do not intend to investigate dynamical and hydrodynamical
interactions between the SMC, the LMC, and the Galaxy:
these interactions will be investigated in detail by Bekki \& Chiba (2008a),
in which the influences of the LMC-SMC interaction
on the LMC's evolution, and those of the LA-Galaxy interaction 
on the evolution of the outer part of the Galactic HI disk
are discussed.

We consider that the best model of the SMC in the present study
needs to reproduce reasonably well the projected locations 
of the MS and the LA in  the Galactic coordinate  $(l,b)$.
Accordingly,  we try to search for the models that  can
reproduce the observed locations of the MS and the LA among models with
different model parameters. 
We first run ``collisionless models'' (labeled as ``C'') 
in which star formation
and gas physics are switched off. 
If we find that a collisionless model
reproduces the locations of the MS and the LA well, 
we then run a ``gas dynamical model'' (labeled as ``G'') 
with model parameters exactly the same as the collisionless model.
Given that the present chemodynamical simulations are numerically costly,
this two-fold investigation is quite 
effective in finding the best model of the SMC among G models.
The LMC is represented by a fully self-consistent disk model
(paper I) 
in some collisionless models so that we can confirm that
the formation of the MS does not depend on whether
the LMC is modeled  as a test particle or a self-consistent disk 
composed of many particles.

 Fig. 3 shows the distribution of ``gas'' particles 
with respect to the center of the Galaxy for  the model C1,
in which orbital evolution and masses of the MCs are consistent
with previous models by GN and C06. This model can  reproduce the observed
location of the MS and the LA reasonably well and shows bifurcated structures
in the MS, which appears to be consistent with observations.
This model is therefore one of the successful models for the MS in the 
present study.
The physical properties of the MS
can be well reproduced by the present models
with different $M_{\rm SMC}$,
{\it as long as we adopt the GN velocity type.}
However, the models with the K06 velocity type
can not reproduce well the observed locations of the MS and the LAs,
irrespective of $M_{\rm SMC}$.
The physical properties of the MS and the LAs in  models
with different velocity types are briefly discussed
in the Appendix A.

Although we run 35 models with different model parameters,
we describe the results for 14 representative models that we regard as important
in terms of the comparison with observations. 
The values of model parameters for 12 models are given in Table 2:
The model numbers (column 1), $M_{\rm SMC}$ (2),
velocity types (3), initial morphologies of the SMC (4),
$R_{\rm SMC}$ (5), 
$r_{\rm s}$ (6), 
$f_{\rm g}$ (7), 
$r_{\rm g}$ (8),
$T_{\rm i}$  (9),
and comments (10).
We mainly describe the results of the ``fiducial model''
in which the simulated properties of the SMC are more
consistent with observations.
Models with K06 velocity types fail to explain
the observed locations of the MS and the 
LA in the galactic coordinate system.
We however  show and discuss  the results of these in order to discuss
the latest proper motion measurements of the MCs (K06, Piatek et al. 2008)
and orbital evolution models of the MCs based on
them (Besla et al. 2007).

We also investigate the ``isolated model'' with no LMC-SMC-Galaxy interaction
in order to confirm that the derived characteristic properties of the SMC
in the G models are due to the LMC-SMC-Galaxy interaction.
We investigate three isolated models (I1,  I2, and I3) with different
$T_{\rm i}$ so that we can confirm which model(s) can be dynamically
stable. We find that the gas disks 
in models with higher $T_{\rm i}$ (e.g., I3
with $T_{\rm i}=10^4$ K) become unrealistically  thick owing to high gaseous
pressure not reasonable for the SMC with a small mass.
We also find that the models with  $T_{\rm i}=250$ K can keep
initially thin gas disks and continue to be stable.
We therefore show the results of  the G models with  $T_{\rm i}=250$ K
in the present study.

\begin{figure}[h]
\begin{center}
\includegraphics[scale=0.8, angle=0]{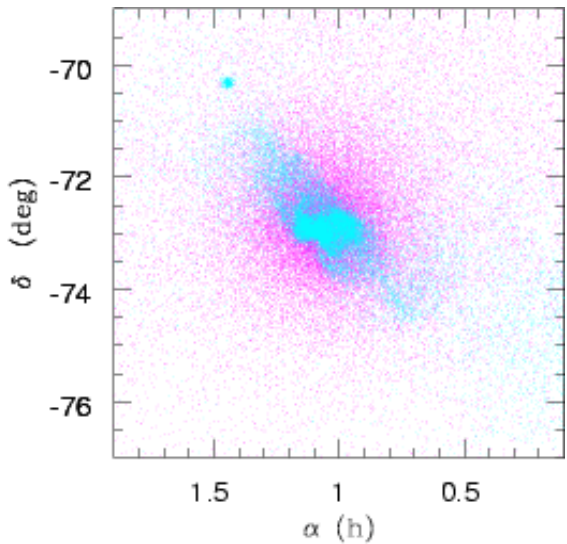}
\caption
{
The final distributions of old stars (magenta) and new ones (cyan)
on the sky (i.e., in the $\alpha-\delta$ coordinate system)
for the fiducial model (G1).
}
\label{figexample}
\end{center}
\end{figure}

\begin{figure}[h]
\begin{center}
\includegraphics[scale=0.4, angle=0]{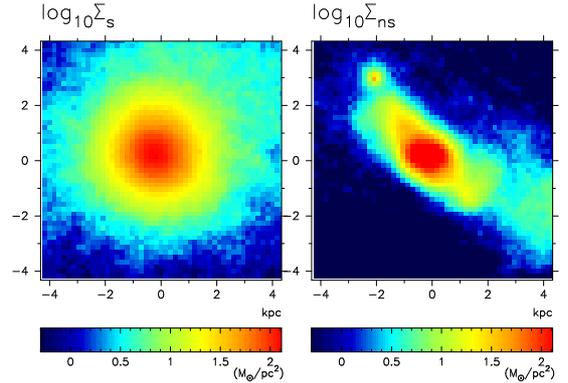}
\caption
{
The final 2D distributions of projected mass densities (${\Sigma}$
in units of ${\rm M}_{\odot}$ pc$^{-2}$) for old
stars (${\Sigma}_{\rm s}$) and new ones (${\Sigma}_{\rm ns}$)
in logarithmic scales for the fiducial model.
Note that the flattened structure can
be clearly seen in this simulated distributions.
}
\label{figexample}
\end{center}
\end{figure}

\section{Results}

\subsection{Fiducial model}

\subsubsection{Dynamical evolution}

Fig. 4 shows how the total masses of dark matter,
old stars, new ones, and gas within the central
5 kpc of  the SMC evolve with time
during strong tidal interaction between the SMC, the LMC, and the Galaxy 
in the fiducial model G1. 
The dark mater, old stellar, and gaseous
components lose 40\%, 41\%, 75\% of their initial masses
during the tidal interaction.
The larger fraction of gas mass lost during the interaction
is due  both  to more efficient tidal stripping of gas
and to gas consumption by star formation.
The central spheroidal component (i.e., old stars)
in this model can not be so strongly influenced
by the tidal fields owing to its initial compactness
so that it does not lose its mass during the strong LMC-SMC
interaction around $T=-1.5$ Gyr.
It however loses  a significant fraction of old stars at
the last LMC-SMC  interaction about 0.2 Gyr ago
so that the MB composed of old stars can be formed.
The final total mass within the central 5 kpc in
this model is $2.4 \times 10^9 {\rm M}_{\odot}$,
which is consistent with the observed total mass of the SMC
(e.g., S04).

Fig. 4 also shows that 
not only  stars but also gas initially in the SMC
can pass through the central 7.5 kpc 
of the LMC (corresponding to the disk size),
in particular, at $T=-1.4$ and $-0.2$ Gyr.
This result confirms the ``Magellanic squall'' (Bekki \& Chiba 2007),
which means efficient mass-transfer from the SMC to the LMC
via the LMC-SMC tidal  interaction.
Fig. 4 furthermore shows that (i)  new stars formed in the SMC
can pass through the LMC and (ii) the total mass of the SMC's new stars
that are located within the LMC at $T=0$ 
is  about $4 \times 10^7 {\rm M}_{\odot}$.
These results imply that metal-poor 
(${\rm [Fe/H]} <-0.6$),  younger stars (or star clusters) with ages less
than 3 Gyr
currently observed in the
halo region of the LMC might well originate from the SMC.
We note that Mu\~noz et al (2006) recently discovered
metal-poor stars with possible metallicities of ${\rm [Fe/H]} = -0.67$ 
in the outer halo of the LMC, though they did not estimate 
ages of these stars.
Owing to the total mass of the LMC being significantly larger than that of
the SMC,  mass transfer from the LMC to the SMC does not occur.

\begin{figure}[h]
\begin{center}
\includegraphics[scale=0.4, angle=0]{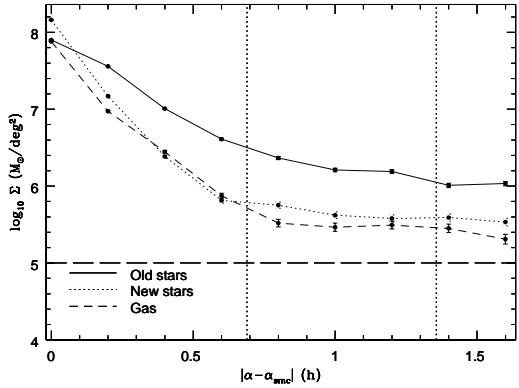}
\caption
{
The final projected mass density profiles ($\Sigma$)
in units of  ${\rm M}_{\odot}$ deg$^{-2}$
along $\alpha$ for
old stars (solid), new ones (dotted), and gas (dashed)
in the fiducial model. The profiles are derived for
for particles  with $\alpha > {\alpha}_{\rm SMC}$,
where ${\alpha}_{\rm SMC}$ is $\alpha$ of the SMC,   so that
the mass profiles in the MB regions can be clearly seen.
For comparison, the MB region where HI and molecular
observations (e.g., Mizuno et al. 2006)
have been done  is shown by two dotted lines.
}
\label{figexample}
\end{center}
\end{figure}

\begin{figure}[h]
\begin{center}
\includegraphics[scale=0.4, angle=0]{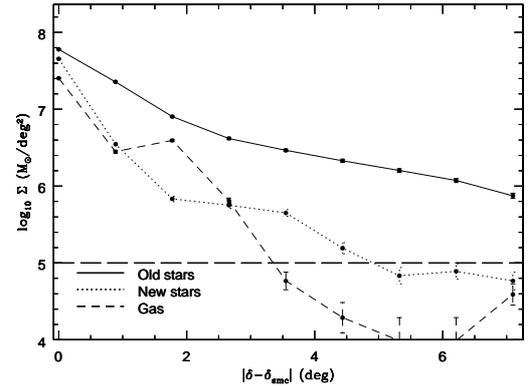}
\caption
{
The same as Fig. 7 but along $\delta$.
The profiles are derived for
for particles  with $\delta > {\delta}_{\rm SMC}$,
where ${\delta}_{\rm SMC}$ is $\delta$ of the SMC,   so that
the observed mass profiles of the stellar halo can be compared
with the simulation.
The physical ranges of $\alpha$ and $\delta$
in Figs. 7 and 8 are the same so that the profiles
along   $\alpha$ and $\delta$ can be compared with each other.
}
\label{figexample}
\end{center}
\end{figure}

The strong tidal interaction can also compress strongly the gas to form
high-density asymmetric structures (e.g., spirals), where star formation
is moderately enhanced (up to $\sim 0.4 {\rm M}_{\odot}$ yr$^{-1}$).
Owing to  the absence of the stellar bar in the dwarf spheroid,
a nuclear starburst is not induced
by the interaction: this result is in a striking contrast with those by 
Noguchi (1988) in which stellar bars formed in the disks of interacting galaxies 
can trigger nuclear starbursts. 
About 30\% of the initial gas is tidally stripped 
to  form the MS,  the LAs, and the MB,
depending on the present locations of the gas.
The stellar spheroid can not be strongly distorted by the tidal interaction
so that 
the final morphology of the SMC shows an old stellar  spheroid
with a gaseous disk more compact than the initial one
and  a compact stellar one composed only of young stars
with ages less  than 2.7 Gyr.

\subsubsection{Stellar structures and kinematics}

As shown in Fig. 5,
the final   projected spatial distribution on  the sky 
(i.e., on  the  $\alpha$-$\delta$ coordinate system) 
are quite different between old and new stars.
New stars originate from the gas disk  disturbed and compressed
by the tidal fields of the LMC and the Galaxy
so that 
the final distribution of new stars is quite elongated:
this structure
is reminiscent of the observed apparently ``barred
structure'' in the SMC.
The old stars, on the other hand,
are not so strongly influenced by the tidal fields as the gas
owing to their initial compact
distribution so that they can  
not be so tidally elongated at $T=0$.
They appear to show an almost spherical shape on the sky,
though the spherical morphology 
is  not so clearly seen
in this plot because of the stripped
stars being  widely distributed in the outer halo region of the SMC. 
These  result clearly explain why  there are significant differences
between the morphology of old stellar populations
(e.g., red giant stars; Harris \& Zarisky 2006) 
and the optical morphology depending strongly on young stellar
populations.
A GC-like compact stellar system  that is formed during
the tidal interaction and thus composed  of new stars can be seen
in the upper left corner of the plot
(i.e.,  $(\alpha,\delta) \approx (1.5{\rm h}, -70.5^{\circ})$).

Fig. 6 shows that the 2D distributions
of the projected surface mass densities
(${\Sigma}$) for  old stars (${\Sigma}_{\rm s}$) 
and for new ones (${\Sigma}_{\rm ns}$)
that  are produced based on the projected distributions of stars
for the region  
shown in Fig. 5.
The old stellar component clearly shows a spherical distribution
on the sky, though  ${\Sigma}_{\rm s}$ of the outer halo  is slightly enhanced
in the northern direction. This north-south asymmetry in the outer stellar halo
is highly likely to be due to the past SMC-LMC-Galaxy  tidal interaction.
The new stellar component is  significantly elongated
and  the major axis of
the distribution in the central region ($R<1$ kpc) is not coincident
with that of the outer part ($R>2$ kpc). 
The lower part appears to be more strongly disturbed
than the upper one
owing to the last LMC-SMC tidal interaction about 0.2 Gyr ago.

\begin{figure}[h]
\begin{center}
\includegraphics[scale=0.6, angle=0]{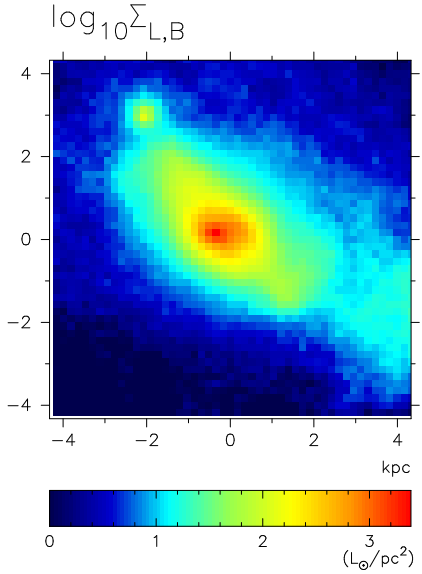}
\caption
{
The final projected 2D distribution of
the  $B$-band luminosity density (${\Sigma}_{\rm L, B}$)
of the SMC for the fiducial model.
The distribution is produced based on the mass distributions
shown in Fig. 6 for adopted stellar population models.
The details of the methods to produce the distribution
are given in the main text.
}
\label{figexample}
\end{center}
\end{figure}

\begin{figure}[h]
\begin{center}
\includegraphics[scale=0.6, angle=0]{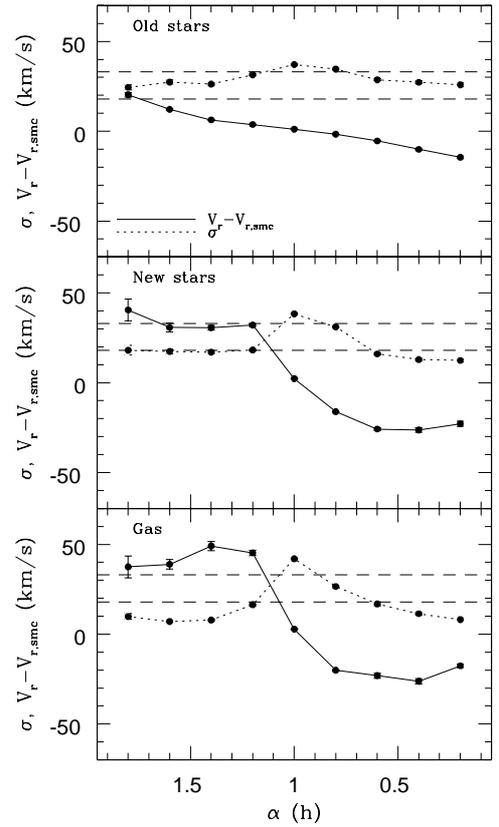}
\caption
{
The final profiles  of line-of-sight velocities
($V_{\rm r}-V_{\rm r,SMC}$, solid) and dispersions ($\sigma$, dotted)
for old stars (top), new ones (middle), and gas (bottom)
in the fiducial  model,
where $V_{\rm r}$ and $V_{\rm r,SMC}$ are the  line-of-sight velocity
of  particles and  of the center of the SMC, respectively.
For comparison, the observed range of the velocity dispersions
of older stars is shown by dashed lines.
}
\label{figexample}
\end{center}
\end{figure}

\begin{figure}[h]
\begin{center}
\includegraphics[scale=0.6, angle=0]{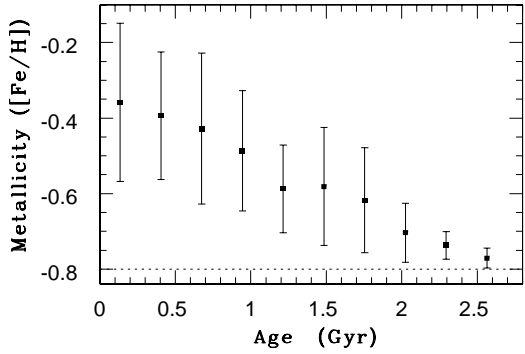}
\caption
{
The AMR of new stars formed for the last $\sim$ 2.7 Gyr evolution
of the SMC in the fiducial model.
The $1\sigma$ dispersions in metallicities are shown
by error bars for each of the age
bins.
}
\label{figexample}
\end{center}
\end{figure}

\begin{figure}[h]
\begin{center}
\includegraphics[scale=0.6, angle=0]{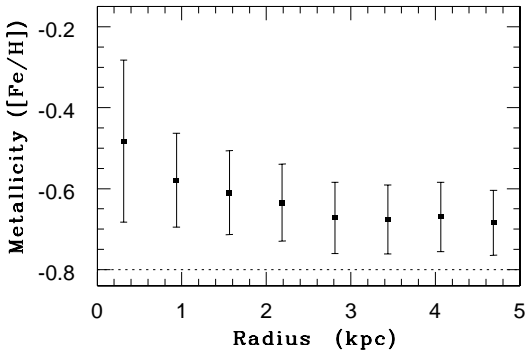}
\caption
{
The radial dependence of the mean metallicities of new stars
in the SMC for the fiducial model. 
The $1\sigma$ dispersions in metallicities are shown
by error bars for each of the radial
bins.
}
\label{figexample}
\end{center}
\end{figure}

\begin{figure}[h]
\begin{center}
\includegraphics[scale=0.4, angle=0]{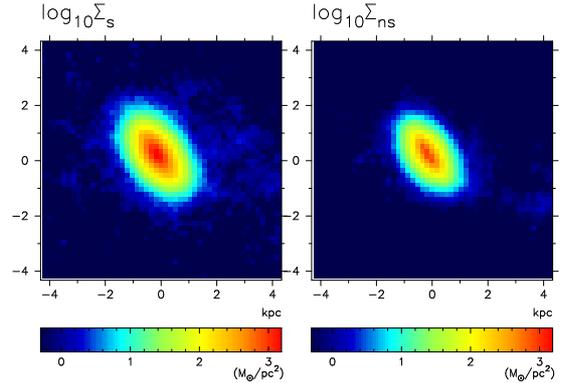}
\caption
{
The same as Fig. 6 but for the disk model G3 in which old stars
are modeled as a rotating disk.
}
\label{figexample}
\end{center}
\end{figure}

\begin{figure}[h]
\begin{center}
\includegraphics[scale=0.6, angle=0]{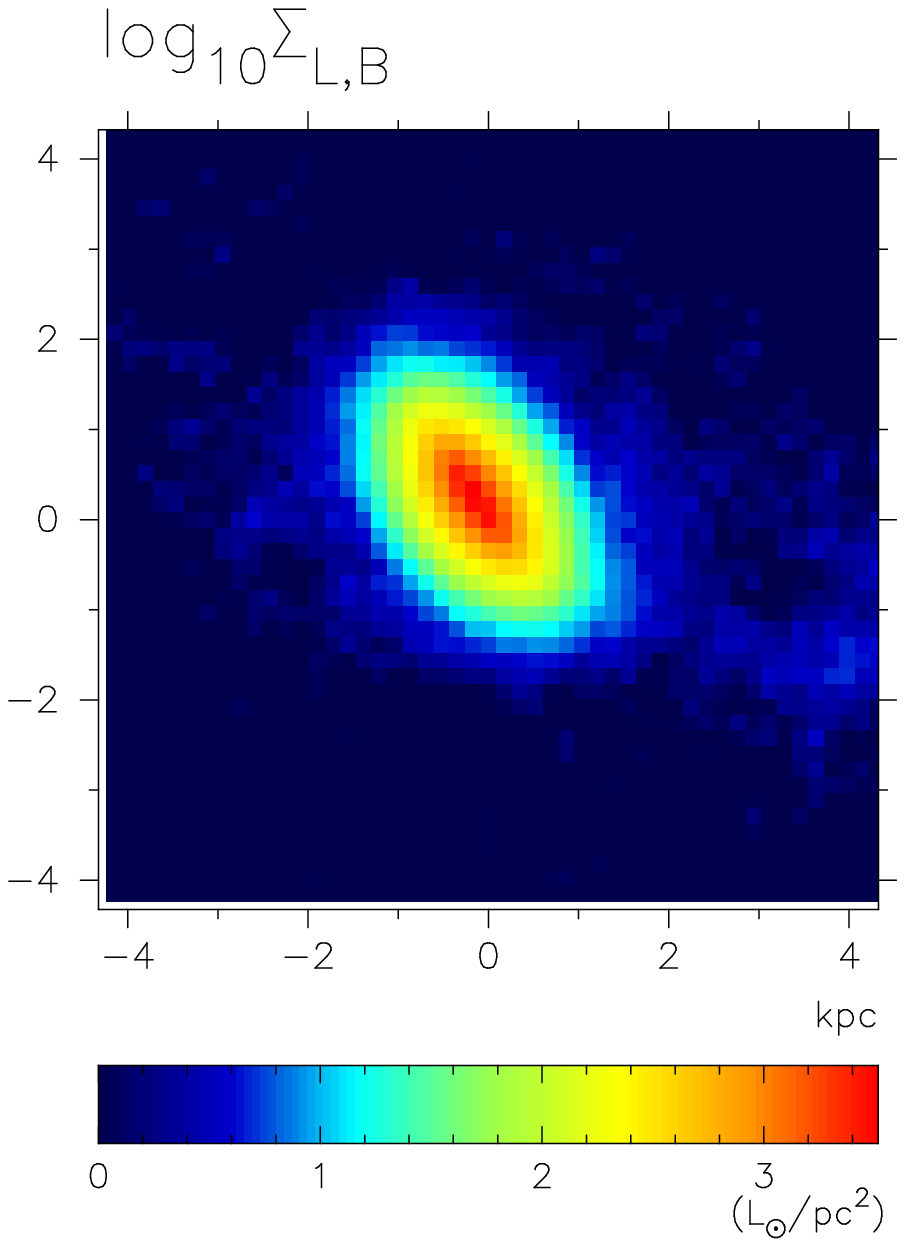}
\caption
{
The same as Fig. 9 but for the model G3.
}
\label{figexample}
\end{center}
\end{figure}

\begin{figure}[h]
\begin{center}
\includegraphics[scale=0.6, angle=0]{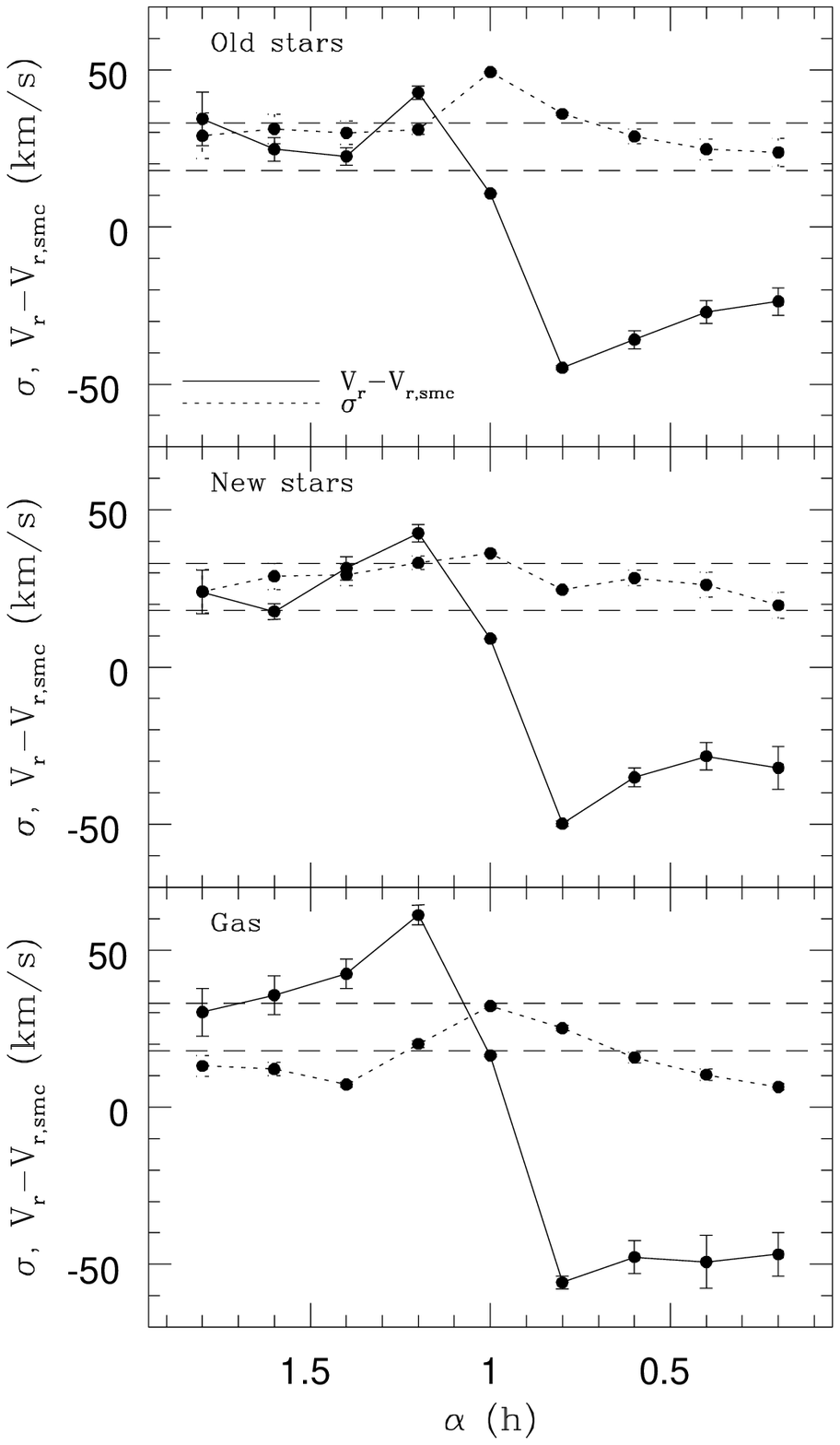}
\caption
{
The same as Fig. 10 but for the model G3.
}
\label{figexample}
\end{center}
\end{figure}

\begin{figure}[h]
\begin{center}
\includegraphics[scale=0.5, angle=0]{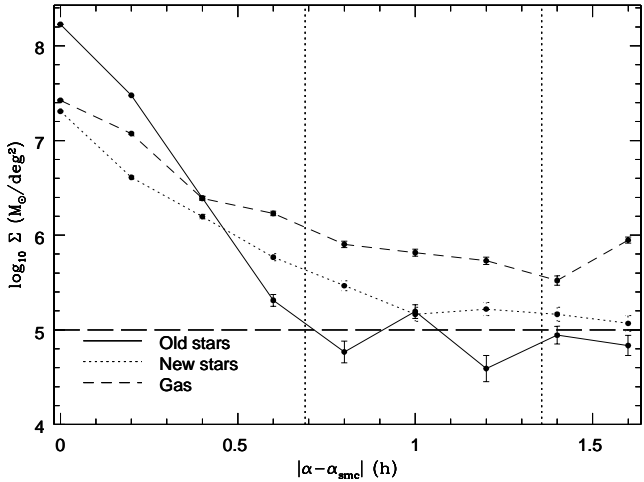}
\caption
{
The same as Fig. 7 but for
the model G2 in which the initial spheroidal component is more compact
than that in G1.
}
\label{figexample}
\end{center}
\end{figure}

\begin{figure}[h]
\begin{center}
\includegraphics[scale=0.6, angle=0]{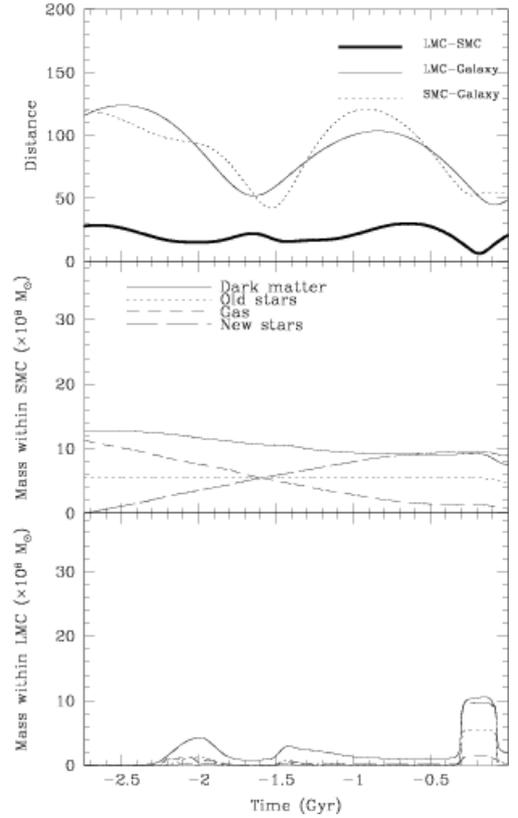}
\caption
{
The same as Fig. 4 but for the model G5 in which
both the initial total mass and the size of the SMC are
significantly smaller than that in G1.
}
\label{figexample}
\end{center}
\end{figure}

\begin{figure}[h]
\begin{center}
\includegraphics[scale=0.6, angle=0]{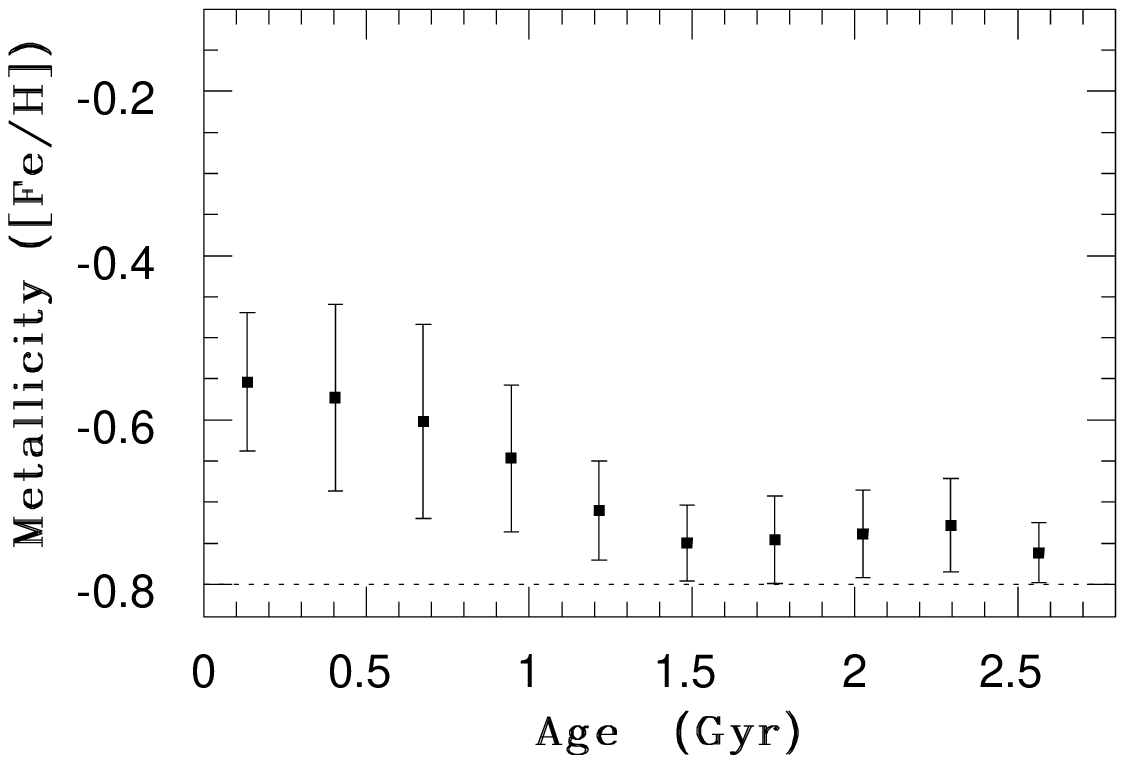}
\caption
{
The same as Fig. 11 but for the model G4 with $y_{\rm met}=0.002$.
}
\label{figexample}
\end{center}
\end{figure}

Owing to the efficient tidal stripping of old 
stars in the last LMC-SMC interaction about 0.2 Gyr ago,
the SMC loses a significant fraction ($\sim 40$\%) of old stars.
As a result of this,  the stripped stars can be widely distributed
between the LMC and the SMC (i.e., the MB region).
Fig. 7 shows that the projected surface mass density (${\Sigma}$)
of old stars  in the MB region
has a relatively flat distribution
with ${\Sigma}$ being significantly higher than $10^6 {\rm M}_{\odot} {\rm deg}^{-2}$.
The presence of this extended halo-like component is due largely to the stripped.
The surface density of new stars in the MB region
is  significantly lower than that of old ones in this model,
though the ${\Sigma}$-ratio of old to new stars in the MB
depends on the initial distributions of old stars in the SMC,
as discussed later in this paper.
As shown in Fig. 7,
there are no significant differences in the  ${\Sigma}$ profiles
between gas and new stars in the MB.

Fig. 8 shows that the  ${\Sigma}$ profiles of old and new stars and gas along  $\delta$
appear to be significantly different from
those along  $\alpha$,
which suggests that the global mass distribution
in the outer part of the SMC  is highly asymmetric owing to the past interaction
with the LMC and the Galaxy.
${\Sigma}$ of gas and new stars along $\alpha$ are systematically higher
that those along $\delta$ for the outer halo region 
($|\alpha-{\alpha}_{\rm SMC}| > 0.7 {\rm h}$ or 
$|\delta-{\delta}_{\rm SMC}| > 3 {\rm deg}$),
which means that a larger fraction of the stripped stars and gas
are located in the MB region.
There are no remarkable differences in ${\Sigma}$ profiles 
along $\alpha$ and $\delta$ for old stars in the SMC.
These results on the possible differences in  ${\Sigma}$ profiles
along  $\alpha$ and  $\delta$ between stellar and gaseous
components will be discussed later in the context
of the formation of the MB and the outer stellar halo in the SMC.

Fig. 9 shows that the 2D $B$-band surface brightness 
(${\Sigma}_{\rm L,B}$) distribution of the SMC
has an irregular shape rather than a regular spherical one,
which is expected from the dominant old stellar populations in this model.
This result clearly suggests that the observed irregular appearance in the SMC
is due largely to the distribution of young stars formed during
the tidal interaction with the LMC and the Galaxy.
However, the simulated  $B$-band morphology is not  so similar
to the observed optical morphology, which has   a ``wing'' and a non-elliptical
appearance in its central region.  
These  apparent differences between the simulated and observed 
morphologies imply that the present models are not so successful
in reproducing the 2D distributions of star-forming regions 
in the SMC.

Fig. 10 shows that the line-of-sight velocity dispersion ($\sigma$) profile
for old stars
has a strong radial gradient with the  maximum $\sigma$ of 37 km s$^{-1}$
and the mean $\sigma$ of 31 km s$^{-1}$. 
The line-of-sight  velocity  profile ($V_{\rm r}-V_{\rm r,SMC}$) for old stars appears  to have
a small amount of rotation ($V < 20$  km s$^{-1}$). 
These simulated  kinematical properties of the old stars
are broadly consistent with the observed ones.
$\sigma$  is  systematically
lower in new stars than in old ones, which reflects that
the new stars originate from gas that can dissipate away the random kinematical
energy produced during the SMC-LMC-Galaxy tidal interaction.
New stars have a significant amount of rotation ($V \sim  40$ km s$^{-1}$)
and thus a high $V/\sigma$ ($ > 1$).
These results suggest that  young stellar populations with ages less than $\sim 3$ Gyr
in the SMC
can have kinematical properties significantly different from older ones
such as PNe, carbon stars, and red giants.

As shown in Fig. 10, the line-of-sight velocity and velocity dispersion
profiles of the gas are clearly different from those of the old stars 
in the sense that the gas show  lower velocity dispersions
($7-42$ km s$^{-1}$),  a higher maximum rotational velocity
($ \sim 49$ km s$^{-1}$), and thus a higher $V/\sigma$ of $\sim 1.2$.
The derived kinematical difference between old stars and gas
and the high maximum rotational velocity are broadly consistent with
the latest kinematical observations by S04.
Fig. 10 also shows that  the velocity profiles for stellar and gaseous components 
are significantly   asymmetric in the sense that the absolute magnitudes
of the velocities appear to be systematically  larger for $\alpha > 1{\rm h}$.
This asymmetry might well be due partly to  the last LMC-SMC interaction
about 0.2 Gyr ago.

\subsubsection{Stellar populations}

In this model with a relatively high $y_{\rm met}=0.004$,
the mean metallicity of the new stars 
evolves from ${\rm [Fe/H]}  = -0.75$ at $T=-2.2$ Gyr 
to  ${\rm [Fe/H]}  = -0.56$ at $T=0$ Gyr
(note that the initial gaseous metallicity is  ${\rm [Fe/H]}  = -0.8$).
It should be stressed here that the final metallicity of new stars
can be significantly smaller than the above value for smaller  $y_{\rm met}$:
the observed mean metallicity of young stars with ages less than $\sim 3$ Gyr
can give constraints on the chemodynamical models of the SMC.
As shown in Fig. 11,
there is an age-metallicity-relation (AMR) of new stars
(i.e., more metal-rich for younger stars),
though the dispersions in stellar metallicities appears to be quite large,
especially  for younger stars.
The shape of the AMR and the metallicity dispersions
also strongly depend on $y_{\rm met}$, as discussed later in this paper.

Fig. 12  shows that the mean metallicities of new stars
are higher in the inner regions than in the outer ones
for the central 3 kpc of the simulated SMC.
This negative metallicity gradient of new stars
reflects the fact that
chemically enriched gas   
due to star formation is transfered to the central region
and converted into new stars there during the SMC-LMC-Galaxy interaction.
The metallicity dispersion is  largest  in the innermost 
radial bin, which implies that new stars currently in the central region
originate from gas that are initially located in  different regions
with different metallicities.
Although the negative metallicity gradient of new stars is
a robust prediction of the present models,
the slope of the gradient and the metallicity dispersions
at radial bins
depend strongly on $y_{\rm met}$.

\subsection{Parameter dependences}

Parameter dependences of stellar and gaseous 
properties are summarized as follows.

(1) Final spatial distributions of  stars do not depend
strongly on model parameters in the dwarf spheroidal  models:
old and new stars have  spherical 
and flattened distributions on the sky, respectively.
The derived structural differences between old and new
stars are broadly consistent with observations,
which implies that the SMC was born as a dwarf spheroidal.

(2) The final structural and kinematical properties
are quite different between the dwarf spheroidal and disk
models. 
Fig. 13 shows  the final
spatial distributions of old and new stars projected onto
the sky for the disk model G3.
The spatial distribution of old stars 
do not appear to be so spherical,
which suggests that dynamical heating by the tidal fields
of the LMC and the Galaxy is not strong enough to 
morphologically transform the initially disky
shape into the  more spherical one.
New stars in the disk model show a bar in the projected
distribution on the sky, which is consistent with the observed
optical morphology of the SMC.
Although only  old stars appear to form the outer extended stellar halo,
the spatial distributions between old and new
stars are quite similar with each other.
Fig. 14 shows that the final $B-$band morphology in the disk model
can be  identified simply as
an inclined  disk galaxy.

(3) The final velocity dispersions 
($\sigma$)
of old stars 
are roughly similar 
($\sigma \sim 30$  km s$^{-1}$) between different dwarf spheroidal models.
The maximum line-of-sight-velocities ($V$)
of old stars
are less than 20 km s$^{-1}$ 
for most of the dwarf spheroidal  models.
It should be here stressed that the apparent non-negligible
amount of rotation ($V<20$ km s$^{-1}$) in old stars
are due to the streaming  motion
of the stars in the outer region of the SMC.
The final velocity dispersions and maximum rotational velocities 
of new stars 
are systematically lower and higher, respectively,
than those of the old ones,
irrespective of model parameters.

(4) There are no significant kinematical differences between
old stars, new ones, and gas in the disk models,
which are inconsistent with observations.
Fig. 15 shows that these three components show higher $V$ 
($\sim 40$ km s$^{-1}$) in the disk model G3
even after the strong tidal heating
of the SMC by the LMC and the Galaxy.
The velocity dispersions of new stars are only slightly
smaller than those of old stars in these disk models.
These results suggest that the disk models are less consistent
with observations than  the dwarf spheroidal ones in terms
of kinematical properties of the SMC.

(5) The surface mass densities ($\Sigma$) in the MB region
depend strongly on initial distributions of stars of the SMC
in the present models.
Fig. 16 shows an example of models in which
$\Sigma$ of old stars can be significantly lower than
those of new ones and gas.
The model G2 shows that $\Sigma$ 
of old stars is lower than $10^5 {\rm M}_{\odot}$ deg$^{-2}$
in some regions of the MB.
This is mainly because a smaller number of old stars
are tidally stripped in this model owing to the more compact
initial distribution of the stars.
These results imply that the observed $\Sigma$
of stars in the MB region can give strong constraints
to the models for the initial distribution of old stars in the SMC.

(6) The SMC can lose a significant fraction (up to  $\sim 40$\%
depending on model parameters)
of its initial mass in the present models. As a result of this,
the final masses of the SMC can be significantly smaller than
the observationally suggested one ($=2.4 \times 10^9 {\rm M}_{\odot}$
within the central 3 kpc of the SMC; S04)
for some models (e.g.,   G6, C1, and C3) with 
$M_{\rm SMC}=3.0 \times 10^9 {\rm M}_{\odot}$:
the initial sizes of the SMC ($R_{\rm SMC}$) need to be
quite small so that too efficient tidal stripping
can be prevented in these low-mass models.
Fig. 17 shows that in the compact, low-mass
model, the SMC loses only 26\% of its initial mass
so that the final mass within the central 5 kpc
($=2.1 \times 10^9 {\rm M}_{\odot}$)
can be similar to  the observed mass of the SMC 
(S04).
The present models with $M_{\rm SMC}=3.0 \times 10^9 {\rm M}_{\odot}$
and $R_{\rm SMC}=7.5$ kpc shows significantly lower final masses
($< 2.0 \times 10^9 {\rm M}_{\odot}$),
which suggests that the previous MS models in C06
may have serious problems in reproducing the present SMC mass.
Fig. 17 confirms that the Magellanic squall (i.e., gas transfer
to the LMC  from the SMC during the LMC-SMC interaction)
can be seen in this low-mass SMC model.

(7) The final mean stellar metallicity,
the AMR of new stars, and the radial stellar
metallicity gradient in the SMC depend
on $y_{\rm met}$.
For example, as shown in Fig. 18,
the AMR in the model G4 with $y_{\rm met}=0.002$
is significantly flatter than that in G1.
Both the stellar metallicity for young stars with
ages less than 0.2 Gyr and the dispersions
in stellar metallicities in this model are significantly
smaller than those  in G1. 
It is, however, currently
unclear which of the two models, G1 and G4,
is more consistent with observations on abundances of young
stars with ages less than 3 Gyrs.

\section{Discussion}

\subsection{Origin of stellar and gaseous kinematics in the SMC}

Previous kinematical studies based on radial velocities 
of older stellar populations such as PNe and red giants
suggested that the stellar component
of the SMC  has the velocity dispersion of
$20-30$ km s$^{-1}$ and appears to have no global rotation
(e.g., Dopita et al. 1995; Suntzeff et al. 1986; Hatzidimitriou et al. 1993).
The latest survey of 2046 red giant stars have suggested
that the stellar component
of the SMC is primarily supported by its velocity dispersion
(Harris \& Zaristky 2006).
Observational studies of HI gas, on the other hand,
showed that the gaseous component has global  rotation with the peak
rotational velocity of about 60 km s$^{-1}$ 
(S04).
No theoretical studies, however,  have yet provided a physical  explanation
for this intriguing kinematical difference
between stellar and gaseous components in the SMC.

The present study has first shown that kinematical differences
between old stars and gas can be clearly seen only in the dwarf
spheroidal models: both stellar and gaseous components of the disk models
show  significant rotation even after strong tidal heating by the LMC
and the Galaxy and  have no kinematical differences between them.
These results imply that the SMC previously did not have
a rotating disk composed both of older stars and gas
but had an extended gas disk kinematically decoupled from
its stellar component.
The rotating gaseous disk is one of essential ingredients 
in the  previous successful  models for the MS (GN, YN, and C06).
Therefore the formation processes of the MS described in these 
studies 
would not be changed significantly if stellar spheroids rather than
stellar disks were  adopted in these studies. 
The previous collisionless  models (BC07b) and the present ones 
in which the SMC is assumed to be a dwarf spheroidal 
can reproduce physical properties of the MS reasonably well
(see Appendix A for more discussions on the MS formation).

The present study suggests that younger stars formed from the rotating
gas disk of the SMC for the last $\sim 2.7$ Gyr can show
a smaller (projected)
central velocity dispersion ($\sigma \sim 20$ km s$^{-1}$)
and a larger (projected) rotational velocity 
($V \sim 40$ km s$^{-1}$).
Although 3D structures of Cepheid variables (possible candidates
of younger stars) along the line-of-sight 
were previously investigated for the SMC (e.g., Caldwell \& Coulson 1986),
kinematical properties of young stellar populations are largely
unknown.
It is thus doubtless worthwhile for
future observations 
to confirm whether or not  younger stellar populations
with ages less than 3 Gyrs in the SMC
show rotational kinematics unlike its older stellar populations.

\subsection{Formation of  a dwarf spheroidal with an extended HI disk}

The present study has first shown that 
the observed dynamical properties of the SMC
can be successfully reproduced by the new dwarf spheroidal
models.
The previous disk models which have been often adopted
in explaining the MS (e.g., YN03) would not be reasonable.
We therefore suggest that the SMC before interacting  strongly
with the LMC and the Galaxy (i.e., more than 3 Gyr ago)
was  a  gas-rich dwarf like
the NGC 6822 with both a spheroidal component
composed of old and intermediate-age stars and an extended HI disk
(e.g., de Blok \&  Walter 2003).
If the SMC was really a gas-rich dwarf spheroidal,
then the following question arises: how were the main body of
the central spheroid  and the outer HI gas disk formed ?

Mayer et al. (2001) showed that 
low-luminosity disk galaxies in the Local Group
can be transformed into dwarf spheroidal and elliptical
galaxies, if they are strongly influenced by tidal fields 
of the giant galaxies like the  Galaxy.
Since this transformation scenario predicts 
quite effective tidal stripping of gas and stars in the outer
parts of disks,
it would not be possible  that the transformed dwarf spheroids
can still contain the outer HI disk.
Therefore the formation of the SMC with an extended HI disk  
would not result from past tidal interaction of the SMC
with the Galaxy.

Recent HI observations of the SMC
have suggested that the total mass of the SMC
derived from the HI rotation  curve
is about $2.4 \times 10^9 {\rm M}_{\odot}$ and thus
consistent with a dynamical model without  a dark matter halo
(e.g., S04).
These observations may well imply that the SMC
is a ``tidal dwarf''  which  can be formed from tidal
tails during strong tidal interaction between two galaxies
(e.g., Duc et al. 2000),
because tidal dwarfs can not contain dark matter components
that are initially in halos of their host galaxies. 
Although there are two luminous galaxies
(i.e., M31 and the Galaxy) in the Local Group
which can be host galaxies for tidal dwarfs,
there  appears to be no observational evidence for 
their past strong interaction.
Thus the tidal dwarf scenario
currently appears to be less promising  as a formation model
of the SMC.

Thus we suggest 
an ``accretion scenario'' in which  the SMC is likely to have developed
its spheroidal component at
the epoch of galaxy formation
and then gradually formed its extended HI disk via
gaseous accretion from its outer halo.
The potential problem of this accretion scenario is that 
the strong tidal field of the LMC and the Galaxy 
could not have allowed
the halo gas of the SMC to be accreted onto  the SMC's disk,
{\it if the SMC has been bound closely to the LMC and the Galaxy since
its formation.}
The present
tidal radius of the present SMC is only $\sim 5.0-7.5$ kpc (GN, YN, and C06),
which implies that only a small amount of gas could have been
accreted onto the SMC without being stripped 
by the LMC and the Galaxy.
The  present distances of the SMC 
with respect to the LMC and the Galaxy may have been 
significantly larger in the past than in the present one.
If this is the case,
the SMC may have developed its massive HI disk through
gradual accretion from the halo to finally become
a dwarf spheroidal with an extended HI disk. 
Other scenarios (e.g., major merging) for the formation of the SMC 
are discussed in detail by Bekki \& Chiba (2008b). 

In the above discussions, we assume that the SMC has a dynamically
hot spheroid with no/little rotation.
It is however possible that the SMC is actually a
rotationally flattened spheroid viewed almost from  face-on
(i.e., along the rotational axis): we can
not detect rotation of the spheroid in the line-of-sight-velocity
field, even if the old stars of the SMC have
rotational kinematics.
If this is the case, the SMC is almost like a polar-ring
galaxy in which the rotational axes of stars and gas are perpendicular
with each other. Although this possibility can not be ruled out
by current observations,  the probability of such a 
configuration of the SMC with respect to the Galaxy 
would be very low. Thus we suggest that the observed apparent
differences in kinematics 
between old stars and HI gas in the SMC reflect
real  differences in {\it internal kinematics} between them:
such differences may well suggest that
the SMC would have experienced accretion/merging event of very gas-rich
objects (e.g., gas-rich dwarfs) in its dynamical history possibly
in a small group where galaxy merging is more likely.

\subsection{On the possible absence of older stars in  the MB}

Observational studies  revealed a number of
intriguing physical properties of stars and gas
in  the inter-Cloud regions
and the MB (e.g., Irwin et al. 1990; Muller et al. 2004;
Mizuno et al. 2006;
Harris 2007; Kato et al. 2007; Nishiyama et al. 2007).
Muller et al. (2004) reported that
the MB is composed of two kinematically and morphologically
distinct structures and shows a remarkable velocity offset
around  $\delta = -73.5^{\circ}$
in the $\delta-V_{\rm rad}$ plane.
The discovery of molecular clouds (Mizuno et al. 2006)
and the possible detection of pre-main-sequence stars (Nishiyama et al. 2007)
in the MB strongly suggest that star formation in the MB is currently ongoing 
and possibly triggered by the last LMC-SMC interaction.
Harris (2007) did not detect an older stellar population
in the MB and thus suggested that the MB originates purely
from gaseous components of the MCs.
Since the observed structural  and kinematical properties
of the MB and the possible physical mechanisms for star formation
in the MB have been already discussed by Muller \& Bekki (2007)
and Bekki \& Chiba (2007), respectively,
we here focus on
the origin of the possible absence of older stellar populations
in the MB.

The present model with the GN velocity type  and 
$M_{\rm smc,s} \sim 10^{9} {\rm M}_{\odot}$ predicts that
the projected surface mass density
in the MB is roughly  
$6 \times 10^4 - 3.0 \times 10^6 {\rm M}_{\odot} {\rm deg}^{-2}$.
The presence of old stars in the MB is due 
largely to the tidal stripping of stars during the last LMC-SMC
close encounter  with its pericenter distance of $\sim 8$ kpc
about 0.2 Gyr ago.
The presence of old stars in the MB thus appears to be inconsistent
with observations by Harris (2007), though the observational upper limit
of the mass density of an older stellar population is not clearly shown
in Harris (2007).
The following two are the possible interpretations 
for this apparent inconsistency between the simulations and the observations.
The first is that the simulated last LMC-SMC interaction needs to be
weaker, so that fewer old stars end up in the MB.
The second is that the initial size of the stellar spheroid (or disk)
in the simulation needs to be smaller to prevent its efficient tidal
stripping by the LMC.

No\"el \& Gallart (2007) have recently found the presence of
intermediate-age and old stars in the southern regions
at 6.5 kpc
from the SMC center.
This result,  which is in a striking contrast to
the results by Harris (2007), suggests that the SMC has an
outer stellar halo component and is possibly more massive
than other observations have ever suggested.
Furthermore, this result  possibly implies that
there could be older stars in the MB,
if the stellar halo of the SMC has a  spherical distribution.
The upper limit for the projected mass density of older 
stars in the MB is crucial to determine whether
the last LMC-SMC interaction is the main mechanism of the MB formation.
We thus suggest that more extensive observational studies
need to be done to provide strong constraints for 
the formation models of the MB.

If the projected surface density of older stars in the MB
is truly  low, as the result by Harris (2007) suggests,
the result by No\"el \& Gallart (2007)
possibly implies  that the stellar halo of the SMC
is quite asymmetric and elongated to the north-south  direction. 
The strong tidal fields of the LMC and the Galaxy
can transform the outer shape of the SMC's stellar halo
so that the recent interaction history, which is determined
by the orbits of the MCs, 
can be imprinted on the shape of the stellar halo.
We thus suggest that
the direction of the major axis of the possibly elongated
stellar halo might well provide some constraints on the orbit
of the SMC around the LMC: the direction would be perpendicular
to the orbit of the SMC and point to the center of the LMC
(thus can constrain the orbital plane of the LMC-SMC system).

\subsection{Dark matter properties of the SMC}

We have first pointed out that (i) the dark matter halo of the SMC is highly likely to have
a very large core ($>5$ kpc) 
and (ii) the observed rotation curve of HI 
and the total mass of the SMC within the central few kpc
can be explained better by the SB profile of the dark matter distribution.
The SB profile with a large core might well explain why the SMC appears to have
no dark matter in the central few kpc (S04), though
a fully self-consistent dynamical model including the dark matter halo
has not been construed yet for the SMC (Bekki \& Chiba 2008a).
If the SMC has a such large, low-density core of its  dark matter halo,
it is highly likely that the halo was initially quite extended 
and therefore lost a substantial amount of its original mass 
owing to the past tidal interaction with the LMC and the Galaxy
since its formation.

As shown in the present simulations,
significant fractions of original masses in the SMC  (up to $\sim$  40\%)
can be lost during the last 2.7 Gyr LMC-SMC-Galaxy interaction.
We have therefore  suggested that 
in order to explain the observed total mass of the SMC
within  the central 3 kpc, 
the original mass of the SMC
should be significantly more massive than 
$M_{\rm SMC}=3.0 \times 10^9 {\rm M}_{\odot}$,
which is adopted in previous MS models (GN and C06).
We have demonstrated that the models with $M_{\rm SMC}=4.5 \times 10^9 {\rm M}_{\odot}$
can better explain the present SMC mass even after the significant mass loss of the SMC
due to the interaction.
It should be noted here that
such models with larger  $M_{\rm SMC}$ and thus larger HI gas mass
can naturally solve
the problems of the previous MS models in explaining the observed total
mass of the MS and the LAs (e.g., YN03). 
We thus suggest that the SMC initially has a larger total mass 
(possibly as much as $4.5 \times 10^9 {\rm M}_{\odot}$)
and a large core in the dark matter distribution.

\section{Conclusions}

We have investigated structural, kinematical, and chemical properties
of stars and gas in the SMC interacting
with the LMC  and the Galaxy based on
chemodynamical simulations.
We have adopted  a new dwarf spheroidal model
with an extended HI gas disk for the SMC
and investigated the chemical and  
dynamical evolution and the star formation
history of the SMC for the last $\sim 2.7$ Gyr.
The main results 
are summarized as follows.

(1) The final
spatial distribution of the old stars projected
onto the sky  is  more  spherical
even after the strong LMC-SMC-Galaxy interaction
whereas that of the new stars is   significantly flattened
and appears to be a bar.
This difference in spatial distributions between the old and new stars
is due to the fact that new stars are formed in the thin  HI disk
of the SMC during the  tidal interaction.
These result suggest that the observed ``bar'' is mostly
composed of new stars recently formed in the SMC.
Some fraction  of old stars initially in the SMC  are stripped to 
form the outer stellar halo of the SMC. 
The distribution of the  stellar halo 
projected onto the sky appears to be slightly asymmetric
in the north-south direction, which possibly reflects
the strong tidal interaction with the LMC about 0.2 Gyr ago.

(2) Old stars have a  line-of-sight   
velocity dispersion ($\sigma$)
of $\sim 30$ km s$^{-1}$ and  weak  rotation 
with the maximum rotational velocity ($V$) of less than 20 km s$^{-1}$
and show asymmetry in the profiles.
New stars have  a smaller $\sigma$
than old ones  and a significant amount of rotation
($V/\sigma >1$).
These kinematical differences between old and new
stars do not depend on model parameters. 
It should be stressed that the derived $V$ of old stars
are highly likely to be overestimated by the stream
motions of the stars along the tidal tails of the SMC.

(3) The HI gas shows
velocity dispersions  of $ \sigma \sim 10-40 $ km s$^{-1}$
depending on the distances from the SMC's center,
a high maximum rotational velocity  ($V \sim 50$ km s$^{-1}$),
and a  spatial distribution similar to that of new stars.
The derived higher velocity dispersion of gas is due to the tidal
disturbance of gas by the LMC and the Galaxy.

(4) The mean metallicity of stars can increase from
${\rm [Fe/H]} \sim -0.80$ to ${\rm [Fe/H]} \sim -0.56$
for $y_{\rm met}=0.004$.
The stellar populations show a weak
AMR in the sense
that younger stars are more likely to be  more metal-rich.
The final stellar [Fe/H] and the AMR depend strongly
on $y_{\rm met}$ 
(e.g., the final metallicity
is ${\rm [Fe/H]} \sim -0.66$ for $y_{\rm met}=0.002$).
More metal-rich stars are more likely to be located
in the inner regions of the SMC, irrespective of model parameters.

(5) Although the MB can be formed in most  models,
the simulated MB
inevitably contains old stars.
This is mainly because some fraction of old stars initially
in the spheroidal component of the SMC
are  tidally stripped to be distributed in the MB region.
The mean surface density of
the stars in the MB 
is different in different models and ranges from 
$ \sim 6 \times 10^4 {\rm M}_{\odot} {\rm deg}^{-2}$
to
$ \sim 3 \times 10^6 {\rm M}_{\odot} {\rm deg}^{-2}$.
If these numbers are inconsistent with observations, 
these results mean that the orbital evolution of the MCs 
for the last 0.2 Gyr in our models
needs to be significantly modified
so as to reduce significantly the stellar mass in the MB.

(6) The disk model shows flattened spatial
distributions and higher maximum rotational velocities ($V>40$ km s$^{-1}$)
in old stars, which appear to be inconsistent with observations.
Thus the dwarf spheroidal model can better explain observations
than the disk one for a given orbit of the SMC
in terms of structures and kinematics of 
old stars in the SMC.
Chemical enrichment processes 
are not so  different between the dwarf and the disk models.

(7)  The simulated MS,
which is composed mostly of HI gas,
shows bifurcation in its  structure and kinematics.
The models based on the GN velocity type  can much better explain
distributions and kinematics of HI gas in the MB than those
based on the K06 one that are derived from the latest
HST proper motion measurements.
Since we did not investigate the MS formation
for all of the possible unbound orbits of the MCs,
it is  unclear whether all of the unbound orbits for the MCs
consistent with  K06 are highly unlikely to explain the MS formation.
These results imply that the physical interpretations
of the HST proper motion measurements need to be reconsidered
in the context of the MS formation.
We  have suggested that future abundance studies of the MS through
QSO absorption lines can provide valuable information on
formation processes and early chemical evolution of the outer
HI disk of the SMC.

(8) The SMC is suggested to initially have
a large core ($>5$ kpc)
in its dark matter distribution
and a larger total mass ($M_{\rm SMC} >3.0 \times 10^9 {\rm M}_{\odot}$)
so that the observed kinematical properties of HI in the present SMC 
can be better explained.
The more massive SMC with a low-density dark matter halo is suggested
to explain better observations of the inner profile of the HI rotation 
curve and of the total gas mass in the MS and the LAs.

The present study adopted the ``classical'' orbits of the MCs
(GN) rather than the new one (K06) in order to compare 
results of the 
present new SMC model with those of the previous ones 
(e.g., Yoshizawa \& Noguchi 2003;  C06) that also adopted
the classical orbits.
We thus have not solved the problems related to the inconsistency
between the predicted velocities of the MCs and the observed
ones (K06). In our forthcoming papers, we discuss how to solve
the problems based on more sophisticated numerical simulations.

\section*{Acknowledgments} 
We are  grateful to the anonymous referee for valuable comments,
which contribute to improve the present paper. 
The numerical simulations reported here were carried out on GRAPE
systems kindly made available by the Astronomical Data Analysis
Center (ADAC) at National Astronomical Observatory of Japan (NAOJ).
K.B. acknowledges financial support from the Australian
Research Council (ARC) and ADAC throughout the course of this work.

\appendix

\begin{figure}[h]
\begin{center}
\includegraphics[scale=0.5, angle=0]{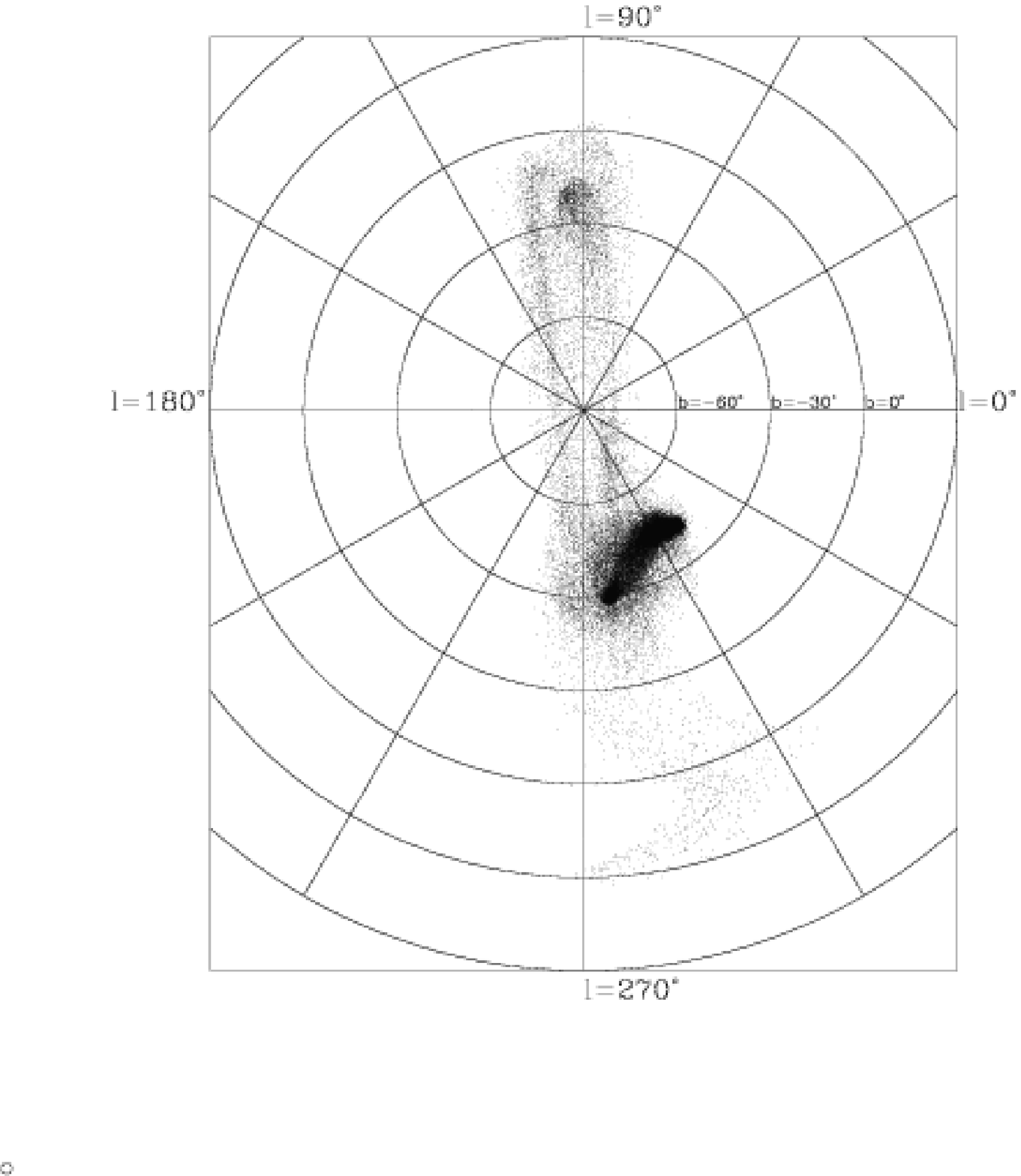}
\caption
{
The final distributions of gas
in the galactic coordinate system $(l.b)$ for the dwarf spheroidal model
C1.
This model is designed to compare with  the early results obtained by
the collisionless MS models of  GN and C06 in which the SMC is modeled
as a
disk
galaxy.
}
\label{figexample}
\end{center}
\end{figure}

\begin{figure}[h]
\begin{center}
\includegraphics[scale=0.45, angle=0]{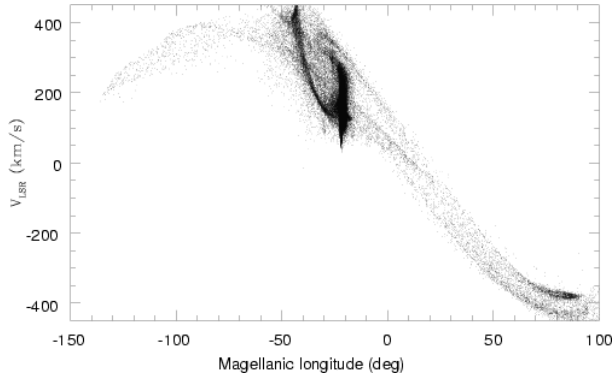}
\caption
{
$V_{\rm LSR}$ (local-standard-of-rest velocities)
of gaseous particles as a function
of Magellanic longitude for the model C1.
}
\label{figexample}
\end{center}
\end{figure}

\begin{figure}[h]
\begin{center}
\includegraphics[scale=0.5, angle=0]{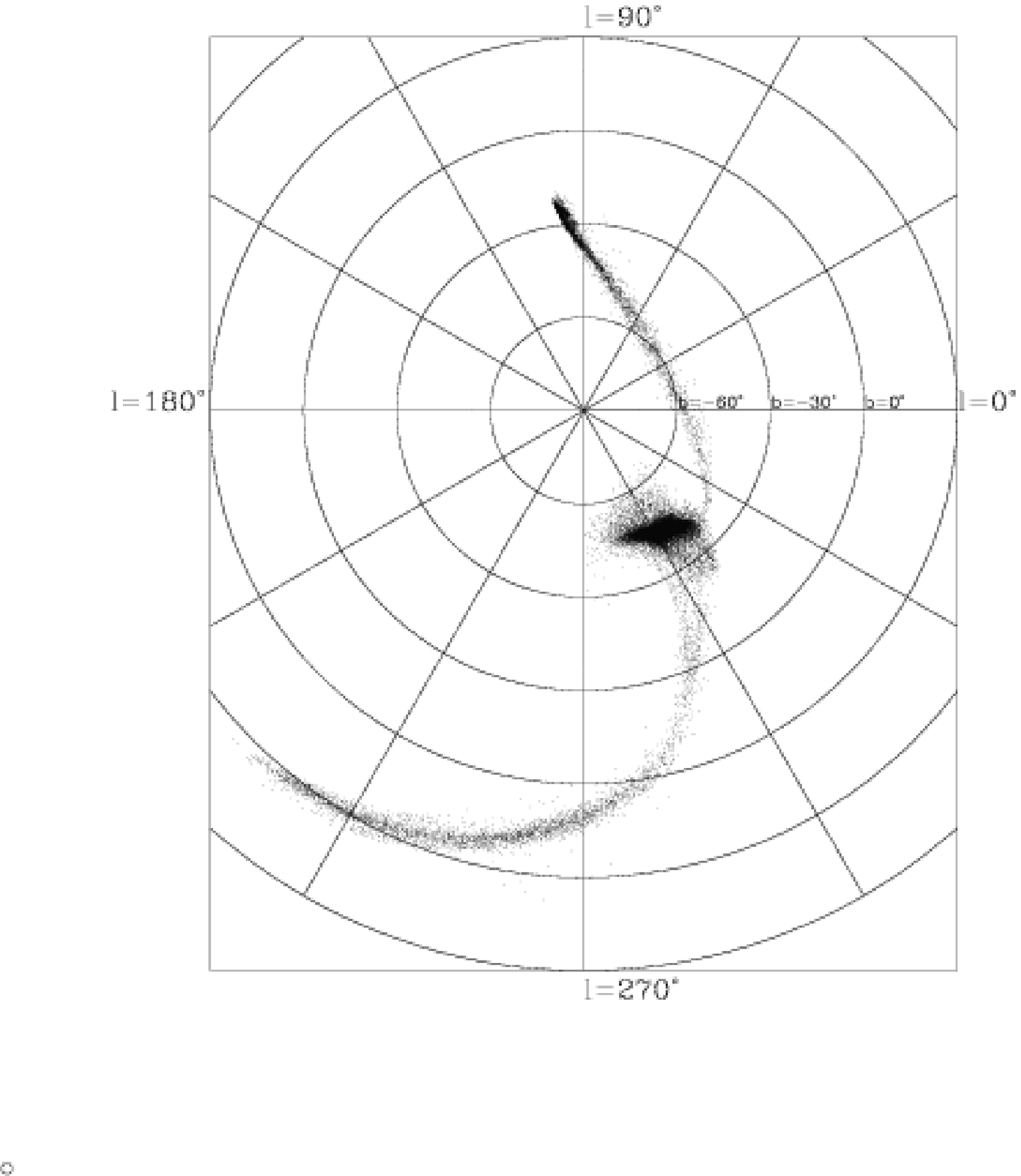}
\caption
{
The same as Fig. 19  but for the
model C2.
}
\label{figexample}
\end{center}
\end{figure}

\section{Formation of the MS}

The previous tidal  models of the MS 
reproduced successfully a number of observed physical
properties of the MS, such as the presence of the LA,
the absence of stars in the MS, and bifurcated structures
and kinematics of the MS (e.g., MF, GN, YN, and C06).
Although the ram pressure models, in which
the MS originates from the gas disk of the LMC,
provided some explanations for the origin of the observed
kinematical properties of the MS,
they failed to explain the presence of the LA
and the bifurcated structure of the MS (e.g., Mastropietro et al. 2005).
Given that the simple version of the tidal  models has some 
difficulties in explaining in the location of the LA on the sky
(Bekki et al. 2008), 
it would be fair to claim that it remains unclear whether
the tidal interaction between the MCs and the Galaxy
is the dominant formation mechanism for the MS. 
Also it is unclear whether the present new dwarf spheroidal  models can
explain physical properties of the MS as well as the previous ones.

Recent observational studies based on proper motion measurements
of the MCs by HST ACS have shed new light on the above unresolved
problem of the MS formation (K06, Besla et al. 2007).
Based on a systematical study of orbital evolution of the MCs,
Besla et al. (2007) suggested that the MCs arrived at the Galaxy
about 0.2 Gyr ago
{\it for the first time} in their histories,
if the observed proper motions of the MCs are correct.
These studies thus imply  that the MS is unlikely to be formed as 
a result of the past tidal interaction of the SMC with the LMC
and the Galaxy about 1.5 Gyr ago.
It is, however, unclear whether theoretical models 
with the orbits of the MCs being consistent with observations by K06
can reproduce the observed properties of the MS.

Thus, we need to investigate both (i) whether
the observed MS can be well reproduced by the present new SMC
models with the GN velocity type
and (ii) whether the present models with the K06 one
can explain the observed properties of the MS as well.
We show the results of the two collisionless  models (C1 and C2),
because other models show similar properties to those of
these two models.
The detailed results of the MS formation in more self-consistent
N-body models for the MCs and the Galaxy
will be described in Bekki \& Chiba (2008a)

Fig. 19 shows that the locations of the MS and the LAs
in the model C1 are broadly consistent with the observed ones
(e.g., Putman et al. 1998), though the shapes of the LAs
appears to be too  broadened.
The bifurcation of the MS can be clearly seen,
which confirm the early results by C06.
Fig. 20 clearly shows the velocity gradient of the MS along
the Magellanic longitudes, which is consistent with observations.
Fig. 20 also shows the bifurcation in kinematics (i.e., $V_{\rm LSR}$)
for the Magellanic longitude larger than 50 degrees.
These simulated bifurcation in the structural and kinematical properties
of the MS will soon  be compared with the corresponding observations
(e.g., Br\"uns  et al. 2005; Nidever et al. 2007)
so that the validity of the models will be addressed.

Fig. 21  shows that the simulated distributions of gas on the sky
are inconsistent with observation in the model C2 with the K06 velocity type,
though two gaseous streams can be formed from the extended gas disk of the SMC
as a result of SMC-Galaxy interaction.
It is also confirmed in the present study that 
the observed locations of the MS and the LA are not reproduced by
the present models with the K06 velocity types irrespective of 
model parameters (e.g., $M_{\rm SMC}$).
There can be mainly two interpretations of the above results.
The first is that the although the 3D space motions of the MCs derived
from the observed proper motions of the MCs are very close
to the true ones, the adopted assumption that the MS
originates from the SMC is wrong.
Besla et al. (2007) proposed a scenario in which  the MS originates from
gaseous components initially within the LMC possibly through
gaseous outflow triggered by stellar feedback effects.
Owing to the lack of numerical studies based on the above scenario,
it remains unclear whether the MS can originate from the LMC, 
if the 3D space motions
of the MCs proposed by K06 are correct.

The second is that the proposed 3D space motions of the MCs by K06
are not close to the true ones so that the inconsistency
in the 3D space motions 
between the tidal models (e.g., GN) and the observationally suggested
ones (K06) can not be a serious problem for the tidal models.
Bekki (2008) first showed  that
the mean proper motion of the SMC derived from the three local fields
can be significantly different from 
that of the center of the mass for  the SMC  so that
the 3D proper motion of the SMC in K06 can deviate significantly
from the true one: {\it interpretations} of the observed proper motion
measurements may or may not be correct. 
These results imply that
the  second interpretation is reasonable, 
though it remains still unclear how close
the 3D space motions adopted in the tidal models
are to the true ones.

Whether the first or the second interpretation is correct,
a number of physical properties of the MS have not been
clearly explained.
In particular,  the ``kink'' in the LA around $b=0^{\circ}$ (C06)
and the ``Magellanic
filaments'' (Nidever et al. 2007)
have not been reproduced by any numerical models of the MS formation
in a self-consistent manner.
As suggested by Bekki et al. (2008),
the origin of the kink around $b=0^{\circ}$ can result
from the hydrodynamical interaction between the outer HI disk
of the Galaxy and the LA.
The apparent physical connections
between the Magellanic filaments and the LMC imply
that there could be some interactions between the MS and the LMC.
We plan to discuss these observations in Bekki \& Chiba (2008a)
based on more sophisticated models in which both a LMC-MS
(or LMC-LA) hydrodynamical interaction and a Galaxy-LA one can be 
self-consistently investigated.

If the MS originates from the SMC, the gaseous abundances
should be very similar to those of the outer gas disk
of the SMC about 1.5 Gyr ago.
The present chemodynamical models imply that
the gaseous metallicity in [Fe/H] ranges from $-0.8$ to $-0.6$,
which is significantly smaller than that of the LMC 
(${\rm [Fe/H]} \sim -0.3$, van den Bergh 2000).
The details of abundance patterns (e.g., [C/Fe], [N/Fe], and [Mg/Fe])
of the MS should be similar to those of the SMC about 1.5 Gyr ago.
We thus suggest that future extensive observational studies
of abundance patterns of gas in  the MS from QSO absorption lines
will provide vital clues to the question as to whether
the MS originate from the SMC or from the LMC.

\end{document}